\documentclass[twocolumn,amsmath,amssymb,aps,pra,groupedaddress]{revtex4-1}

\usepackage{graphicx}% Include figure files
\usepackage{dcolumn}% Align table columns on decimal point
\usepackage{bm}% bold math
\usepackage{color, soul}

\usepackage[table]{xcolor}
\usepackage{wasysym}
\usepackage{subfigure}
\usepackage{commath}
\usepackage{braket}
\usepackage{amsmath}
\let\oldhat\hat
\renewcommand{\hat}[1]{\oldhat{\mathbf{#1}}}
\newcommand{\vect}[1]{\boldsymbol{#1}} % Uncomment for BOLD vectors

\usepackage{amsmath}
\usepackage{amsfonts}
\usepackage{amssymb}
\usepackage{amsthm}

\usepackage{siunitx}
\usepackage{booktabs}
\usepackage{multirow}
\usepackage{epstopdf}
\usepackage{longtable}
\usepackage{array}
\usepackage{latexsym}
\usepackage[latin1]{inputenc}
\usepackage{textcomp}
\usepackage[british]{babel}

\begin{document}

\preprint{}%AIP/123-QED}

\title{Coherent population trapping with polarization modulation}

\author{Peter Yun}
\email{enxue.yun@obspm.fr}
\affiliation{LNE-SYRTE, Observatoire de Paris, PSL Research University, CNRS, Sorbonne Universits, UPMC Univ. Paris 06, 61 Avenue de l'Observatoire, 75014 Paris, France}

\author{St\'ephane Gu\'erandel}
\affiliation{LNE-SYRTE, Observatoire de Paris, PSL Research University, CNRS, Sorbonne Universits, UPMC Univ. Paris 06, 61 Avenue de l'Observatoire, 75014 Paris, France}

\author{Emeric de Clercq}
\affiliation{LNE-SYRTE, Observatoire de Paris, PSL Research University, CNRS, Sorbonne Universits, UPMC Univ. Paris 06, 61 Avenue de l'Observatoire, 75014 Paris, France}

\date{\today}

\begin{abstract}
Coherent population trapping (CPT) is extensively studied for future vapor cell clocks of high frequency stability.
In the constructive polarization modulation CPT scheme, a bichromatic laser field with polarization and phase synchronously modulated is applied on an atomic medium. A high contrast CPT signal is observed in this so-called double-modulation configuration, due to the fact that the atomic population does not leak to the extreme Zeeman states, and that the two CPT dark states, which are produced successively by the alternate polarizations, add constructively.
Here we experimentally investigate CPT signal dynamics first in the usual configuration, a single circular polarization. The double-modulation scheme is then addressed in both cases: one pulse Rabi interaction and two pulses Ramsey interaction. The impact and the optimization of the experimental parameters involved in the time sequence are reviewed.
We show that a simple seven-level model explains the experimental observations.
The double-modulation scheme yields a high contrast similar to the one of other high contrast configurations like push-pull optical pumping or crossed linear polarization scheme, with a setup allowing a higher compactness. The constructive polarization modulation is attractive for atomic clock, atomic magnetometer and high precision spectroscopy applications.
\end{abstract}

\maketitle
\section{\label{sec:level1}Introduction}

Atomic states preparation is required for many quantum optics experiments.
Among various methods, light is a precise and effective tool to manipulate the atom into a desired state.
For instance, circularly polarized resonant light is used to optically pump the atoms into an extreme ground state Zeeman sub-level to create a macroscopic atomic polarization, a crucial step in spin-exchange optical pumping experiments \cite{Happer97}. A circularly polarized bichromatic laser is also commonly employed in coherent population trapping (CPT) experiments on alkali-metal vapor  \cite{Wynands99, Vanier05}, in which a coherent two-frequency laser beam couples two ground states to a common excited state forming a ``$\Lambda$" shaped system. When the difference of the two frequencies equals the ground state splitting, i.e. at Raman resonance, the atoms are trapped into a state superposition, the dark state or CPT state, which is decoupled from the laser field \cite{Arimondo96}. Atoms trapped in a dark state do not absorb the laser light, so that the light intensity transmitted by the medium increases. As the CPT resonance linewidth is defined by the coherence lifetime in the ground state, a narrow resonance is observed which can be used to make compact atomic clocks or magnetometers.

Atomic clocks using alkali-metal atoms \cite{Vanier05, Shah10} are based on the microwave transition $\ket{F,m_F=0}-\ket{F+1,m_F=0}$, the ($0,0$) transition, because of its reduced sensitivity to the magnetic field. Here  $F$ and $F+1$ are the total angular momenta of the hyperfine levels of the ground state, $m_F$ is the magnetic quantum number. Realization of a superposition state  of the $m_F=0$ states requires a circularly polarized field because of selection rules  (the transition  $\ket{g,F,m_F=0}-\ket{e,F,m_F=0}$ is forbidden, $g$ and $e$ hold for the ground and the excited state respectively). But in practice, the Zeeman sublevels of the ground state complicate the situation.
A circularly polarized bichromatic laser will  pump the atomic populations into the ``end-states", corresponding to the Zeeman sub-levels of maximum or minimum magnetic quantum number $m_F$. This Zeeman optical pumping \cite{Happer72},
not only decreases the population involved in the ($0,0$)  CPT resonance, reducing the maximum transmitted power,
but also increases the transmitted laser power off Raman resonance, the background level.

This drawback might be overcome by using counter-rotating circular polarizations. However the dark states built by these two polarizations are out of phase in the involved double lambda system (``$\Lambda \Lambda$"), because one of the four electric dipole moments involved is of opposite sign to the three others.
In the rotating frame and at resonance, the dark states built with the clock states $\ket{1},\, \ket{3}$ (see Fig.~\ref{fig2}) with $\sigma^+$ and $\sigma^-$ polarizations can be written \cite{ Taichenachev06,Yun14}:
\begin{eqnarray}
\left\lbrace
\begin{split}
\ket{d_+}&=\Omega_3\ket{1}- \Omega_1e^{i\phi_+} \ket{3},\\
\ket{d_-}&= \Omega_3\ket{1}+ \Omega_1e^{i\phi_-}\ket{3}.\\
\end{split}
\right.
\label{darkstates}
\end{eqnarray}
where $\Omega_{i}$ is the Rabi frequency of the laser field tuned to the $\ket{i}-\ket{e}$ transition, $\phi_+(\phi_-)$ is the phase difference between both fields in  $\sigma^+(\sigma^-)$ polarization.
So that, with the same phase difference, the dark state built by a circular polarization
is destroyed by the counter-rotating polarization, and vice versa. This is why the clock transition cannot be observed with lin$\parallel$lin polarizations \cite{Zibrov06, Taichenachev06}. It has been shown that counter-rotating circular
polarizations are efficient if one of the two circularly polarized beam is phase
delayed by $\pi$ relative to the other \cite{Jau04, Shah06,Liu13},
or equivalently by linearly and orthogonally polarized beams \cite{Zanon05b}.

Another possibility is to use a beam with alternating opposite circular polarizations \cite{Huang12}. The phase shift of both dark states can be compensated by properly choosing the phase difference between the two components of the bichromatic laser, the Raman phase \cite{Yun14}. We have shown that if the two components of the bichromatic field are in phase for one polarization, they must be out of phase of $\pi$ in the opposite polarization \cite{Yun14}.
Such a simultaneous modulation of polarization and Raman phase
prevents the atomic population from being pumped into the end Zeeman states.
Atomic population accumulates every cycle in the common dark state whereas a strong absorption remains off-resonance. Both effects would enhance the CPT resonance amplitude, which is one of the most important parameters in atomic clock, atomic magnetometer and high precision spectroscopy applications.

The paper is divided into several sections and a conclusion. Section II describes the experimental setup, section III presents a 7-level theoretical model based on Bloch equations interacting with a near-resonant double-modulated bichromatic laser field. The optical thickness effect is also taken into consideration. This model correctly reveals most aspects of the real system and matches well the experimental results. Section IV presents experimental results. We experimentally study the build-up and decrease of the CPT transitions amplitudes in presence of a single circularly polarized beam. The double modulation (polarization and phase) scheme is then addressed. We focus on the effect of the modulation frequency. The last section of this part addresses the double modulation technique with
two pulses (Ramsey-type) interrogation.

\section{\label{sec:level1}Experimental setup}

\begin{figure}[htb]
\centering
\includegraphics[width=0.90\linewidth]{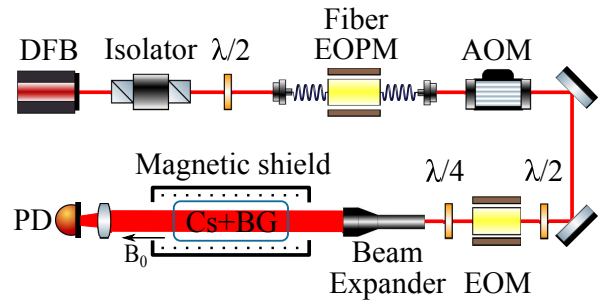}
\caption{(Color online) Setup for double modulation CPT experiments. DFB: diode laser, EOPM: phase modulator, AOM: acousto-optic modulator, EOM: polarization modulator, BG: buffer gas, PD: photodiode.}
\label{fig1}
\end{figure}

Our experimental setup depicted in Fig.~\ref{fig1} is basically the same as the one
described in our previous studies  \cite{Yun14}.
The main difference lies in that an electro-optical phase modulator (EOPM, Photline: NIR-MPX$850$-LN$08$)
rather than a Mach-Zehnder modulator (MZM) is used to generate optical sidebands.
We found it adds less noise to the output laser beam intensity than the MZM.
Unlike the MZM, in which the bichromatic laser beam shares $98.5\%$ of the total laser power,
the ratio is only $68\%$ in the EOPM output.

A monochromatic laser beam emitted by a distributed feedback (DFB) diode laser is tuned to the $D_1$ line of cesium (894\,\si{\nano\meter}).
 The EOPM, driven at  $4.596$\,\si{\giga\hertz}  with $26$\,dBm microwave power, generates phase-coherent optical sidebands.
 The first sidebands ($\pm 1$) are separated by \SI{9.192}{\giga\hertz}, i.e., the hyperfine splitting frequency of the Cs ground state.
 They are tuned to the Cs transitions $\ket{6^2S_{1/2}, F=3}\rightarrow\ket{6^2P_{1/2}, F'=4}$ and $\ket{6^2S_{1/2}, F=4}\rightarrow\ket{6^2P_{1/2}, F'=4}$ respectively, where $F$ ($F'$) is the hyperfine quantum number of the ground (excited) state. A ($0$-$\pi$/$2$) modulation of the phase of the driving microwave signal allows a ($0-\pi$) modulation of the phase between these two sidebands, the Raman phase.
After passing through a polarization modulator, which is synchronously modulated with the Raman phase, we get a double-modulated bichromatic laser. The polarization modulator is formed by an electro-optic amplitude modulator (EOM, Thorlabs: EO-AM-NR-C$2$), used as a voltage tunable half-wave plate, with a $18$\,dB extinction ratio and a rising and falling time $t_{EOM}$ of about $2.5$\,\si{\micro\second}. The beam polarization modulation between two linear crossed polarizations at the output of the polarization modulator is converted into a modulation between opposite circular polarizations by a quarter-wave plate. The doubly modulated Raman laser beam is then expanded to a diameter of $10$\,\si{\milli\meter} before  the vapor cell. The cylindrical Cs vapor cell, \SI{20}{\milli\meter} in diameter and \SI{50}{\milli\meter} long, is filled with $21$\,torr of buffer gas (mixture argon-nitrogen). The cell temperature is stabilized at \SI{32.5}{\degreeCelsius}. A solenoid coil surrounded by two magnetic shields provides a uniform magnetic field of \SI{5.16}{\micro\tesla} in order to remove the Zeeman degeneracy.
To study one pulse (Rabi) and two pulses (Ramsey) CPT spectrum, we add an acousto-optic modulator (AOM) used as an optical switch before the polarization modulator.

\section{\label{sec:level1}Theoretical model}
\subsection{\label{sec:level2}Double-modulation laser field}

The electric field of the monochromatic laser source can be represented by: $\vect{E_1}=\varepsilon  \cos(\omega_0t)\hat{e}_1$. Here we assume the linear polarized basis unit vector $\hat{e}_1=(\hat{e}_x+\hat{e}_y)/\sqrt{2}$.
With the help of an EOPM driven by a microwave signal (frequency $\omega_m$, modulation index $m$) the laser electric field becomes:
$\vect{E_2}=\varepsilon  \cos(\omega_0t+m \sin(\omega_m t+\phi))\hat{e}_1$.
The first sidebands yield a bichromatic laser field:
\begin{eqnarray}
   \vect{E_3}=\varepsilon_0 [\cos(\omega_+t+\phi)-\cos(\omega_-t-\phi])\hat{e}_1.
 \end{eqnarray}
Here $\omega_\pm=\omega_0 \pm \omega_m$, $\varepsilon_0=\varepsilon J_1(m)$ and  $J_1$ is the Bessel function of the first kind of order $1$.
The microwave phase $\phi$ is further modulated by a square wave $f_1(t)$, which switches
between $0$ and $1$, with amplitude $\phi_m$, i.e., $\phi=\phi_m f_1(t)$.

This phase-modulated bichromatic laser field is also polarization modulated to realize a double-modulated field. The polarization modulator is based on an EOM driven by the half-wavelength voltage with a square wave modulation $f_2(t)$.
With the light polarization vector $\hat{e}_1$ oriented at $\ang{45}$ with respect
to the EOM crystal principal axes, x and y, the output field can be written as:

\begin{eqnarray}
  &\vect{E_4}&=\frac{\varepsilon_0}{\sqrt{2}}\times\nonumber\\
  \{&[&\cos(\omega_+t+\phi+\phi_{x+})-\cos(\omega_-t-\phi+\phi_{x-})]\hat{e}_x \nonumber\\
  +&[&\cos(\omega_+t+\phi+\phi_{y+})-\cos(\omega_-t-\phi+\phi_{y-})]\hat{e}_y \}\nonumber.\\
 \end{eqnarray}
The phases induced by the EOM crystal are: $\phi_{x\pm}=(2\pi l_0 n_{x\pm})/\lambda_\pm$,
$\phi_{y\pm}=(2\pi l_0 n_{y\pm})/\lambda_\pm$,
here $l_0$ is the crystal length, $\lambda_\pm=2\pi c/\omega_\pm$,  $n_{x\pm}$ and $n_{y\pm}$ are the refractive indices of ordinary and extraordinary lights at wavelength of $\lambda_\pm$, respectively.
As $n_{x(y)+}\approx n_{x(y)-}$ and $\abs{\lambda_+-\lambda_-}/\lambda_0=2.7\times10^{-5}$,
with $\lambda_0=(\lambda_++\lambda_-)/2$, the difference of the phase delays between $\omega_+$  and $\omega_-$ are negligible, i.e., $\phi_{x+}\approx\phi_{x-}\equiv\phi_{x}$, $\phi_{y+}\approx\phi_{y-}\equiv\phi_{y}$.
Since $\phi_y- \phi_x=2\pi l_0 (n_y-n_x)/\lambda_0 \propto V_{pm}$,
a proper voltage $V_{pm}$ is applied to the EOM crystal to match
the condition: $\phi_x= \phi_y+2\varphi$, $\varphi=f_2(t)\pi/2$, here $f_2(t)$ is
the same square wave as $f_1(t)$, but may have a different response time. We obtain:
\begin{eqnarray}
\vect{E_5}=\varepsilon_0\{&\cos(\varphi)&[\cos(\omega_+t+\varphi+\phi)-\cos(\omega_-t+\varphi-\phi)]\hat{e}_1\nonumber\\
                         -&\sin(\varphi)&[\sin(\omega_+t+\varphi+\phi)-\sin(\omega_-t+\varphi-\phi)]\hat{e}_2\}. \nonumber\\
 \end{eqnarray}
Here $\hat{e}_2=(\hat{e}_x-\hat{e}_y)/\sqrt{2}$ and we have dropped the common phase $\phi_y$.
Finally with a quarter-wave plate, we get the double-modulated laser beam:

\begin{eqnarray}
 \vect{E_6}=\frac{\varepsilon_0}{2}\lbrace &\cos(\varphi)&[i(e^{i(\omega_+t+\varphi+\phi)}-e^{i(\omega_-t+\varphi-\phi)})\hat{e}_- +C.C.]  \nonumber\\                                   +&\sin(\varphi)&[(e^{i(\omega_+t+\varphi+\phi)}-e^{i(\omega_-t+\varphi-\phi)})\hat{e}_+ + C.C.] \rbrace. \nonumber\\
 \end{eqnarray}
with $\hat{e}_\pm=\mp(\hat{e}_x\pm i\hat{e}_y)/\sqrt{2}$.

\subsection{\label{sec:level2}Maxwell-Liouville equations}

\begin{figure}[htb]
\centering
\includegraphics[width=0.60\linewidth]{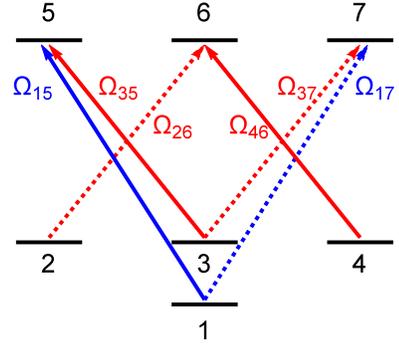}
\caption{(Color online) The 7-level system coupled with a double-modulated bichromatic laser.}
\label{fig2}
\end{figure}

To simplify the problem, a seven-level system (Fig.~\ref{fig2}) is adopted to represent the $25$ involved levels of Cs.
The evolution of  the density matrix for atoms at rest is given by the Liouville equation,
\begin{eqnarray}
  \dot{\rho}=-\frac{i}{\hbar}[H,\rho]-R\rho.
  \label{Hamiltonian}
 \end{eqnarray}
where $H=H_0+V$, $H_0=\hbar \sum_{i=1}^{7}\omega_i\ket{i}\bra{i}$ is the Hamiltonian of the unperturbed atom, $R$ is the relaxation matrix.
The Zeeman effect is neglected. The interaction Hamiltonian is $V=-\vect{D} \cdot \vect{E_6}$, with $\vect{D}$ the electric dipole operator. As the light is polarized $\sigma^+$ or $\sigma^-$, but not both in the same time, Zeeman coherences are not build, and we do not take them into account.
We use the following transformation of the off-diagonal coherences:
\begin{eqnarray}
\rho_{15}&=&\tilde\rho_{15}\exp(i\omega_+t), \nonumber\\
\rho_{35}&=&\tilde\rho_{35}\exp(i\omega_-t),\, \rho_{46}=\tilde\rho_{46}\exp(i\omega_-t), \nonumber\\
\rho_{17}&=&\tilde\rho_{17}\exp(i\omega_+t), \nonumber\\
\rho_{26}&=&\tilde\rho_{26}\exp(i\omega_-t),\, \rho_{37}=\tilde\rho_{37}\exp(i\omega_-t), \nonumber\\
\rho_{13}&=&\tilde\rho_{13}\exp(i(\omega_+-\omega_-)t).
\label{tilderho} \end{eqnarray}
and $\rho_{ji}=\rho_{ij}^*,\, \tilde\rho_{ji}=\tilde\rho_{ij}^*$.
Within the secular approximation, the interaction Hamiltonian becomes:

\begin{eqnarray}
V=&-&\frac{\hbar}{2}[\Omega_{15}e^{i(\varphi+\phi+\pi/2)}\ket{1}\bra{5}+\Omega_{17}e^{i(\varphi+\phi)}\ket{1}\bra{7}\nonumber\\
&-&e^{i(\varphi-\phi+\pi/2)}(\Omega_{35}\ket{3}\bra{5}+\Omega_{46}\ket{4}\bra{6})\nonumber\\
&-&e^{i(\varphi-\phi)}(\Omega_{37}\ket{3}\bra{7}+\Omega_{26}\ket{2}\bra{6})+H.C.].
    \label{V}
 \end{eqnarray}

The Rabi frequency $\Omega_{ij}$ ($=\Omega_{ji}$) are defined as:
\begin{eqnarray}
&&\Omega_{15}=\cos(\varphi)\Omega,\nonumber\\
&&\Omega_{35}=\Omega_{46}= \cos(\varphi)\Omega, \nonumber\\
&&\Omega_{17}=\sin(\varphi)\Omega, \nonumber\\
&&\Omega_{37}=\Omega_{26}= -\sin(\varphi)\Omega.
\label{Omega}
\end{eqnarray}
with $\Omega=d\varepsilon_0/\hbar$, and we assume $\Omega$ to be a real number. Here $d$ is the electric dipole matrix element of the transition.
For the Cs $D_1$ line \cite{Steck}, the involved matrix elements are such that:
\begin{eqnarray}
d_{17}=d_{15}=d_{35}=d_{46}=-d_{37}=-d_{26}=d.
 \end{eqnarray}

\setlength{\tabcolsep}{10pt}
\renewcommand{\arraystretch}{1.2}

\begin{table}[htb]
\caption{\label{tab:table1} $V_{ij}$ terms (in unit of $\hbar/2$) in SM ($\phi_m=0$) and DM ($\phi_m=\pi/2$) cases.}
%\rowcolors{1}{red}{red}
  \begin{tabular}{|c|c|c|c|c|}
  \hline
    %\toprule
   \textrm{$\phi_m$}  &\multicolumn{2}{c|}{0} &\multicolumn{2}{c|}{$\pi$/2}\\

      %\midrule
      \colrule
   $ f(t)$            & 1        & 0          & 1          & 0  \\
      \colrule
   $ V_{15}$     & 0             & $-i\Omega$ & 0          & $-i\Omega$       \\
   $ V_{35(46)}$ & 0             & $+i\Omega$ & 0          & $+i\Omega$       \\
   $ V_{17} $    & $-i\Omega$    & 0          & $+\Omega$   & 0 \\
   $ V_{37(26)}$ & $-i\Omega$    & 0          & $-\Omega$  & 0  \\
    \hline
   % \bottomrule
  \end{tabular}
  \label{table1}
\end{table}

In Table \ref{table1}, we specify the Rabi frequencies in the polarization modulation (SM) and double-modulation (DM) cases.
As we have already shown in ref. \onlinecite{Yun14}, the value of
$S\equiv\frac{V_{15}}{V_{35}}/\frac{V_{17}}{V_{37}}$ will lead to the constructive or destructive effect of the opposite polarizations on the two dark states.
In the SM case, the phase shift  of $V_{37}$ with respect to the three others ($V_{15}$, $V_{35}$ and $V_{17}$) in the ``$\Lambda\Lambda$" system leads to $S=-1$, thus the two dark states are mutually destructive. While in the DM case, the additional Raman phase modulation offers the compensation to meet $S=1$, leading thus to a common dark state.

In order to compare with the actual experimental system, we have found it is necessary to take into account the absorption along the vapor cell length.
Similarly to the procedure described in ref.~\onlinecite{Godone02}, we get the Maxwell-Liouville equation sets
~(\ref{Maxwell-Liouville-2}3) shown in Appendix. The model takes into account the following relaxation rates. $\Gamma$ is the decay rate of the populations of the excited levels, it includes the effects of spontaneous emission and collisions with buffer gas atoms and molecules. When an optical transition between the levels $\ket{i}$ and $\ket{j}$ is allowed, the corresponding decay rate is assumed to be $\Gamma_{ij}=\Gamma/3$.
The transition $\ket{6^2S_{1/2}, F=4, m=0}\rightarrow\ket{6^2P_{1/2}, F'=4, m=0}$, levels $3$ and $6$ in Fig.~\ref{fig2}, is forbidden by the selection rules, $\Gamma_{36} =0$. Optical coherences are damped with a rate $\Gamma/2$.
$\gamma_1$ is the relaxation rate of the differences of population between ground-state hyperfine levels towards the thermal equilibrium,
which is the equipartition of the population. $\gamma_2$ is the relaxation rate of the hyperfine coherence, $\rho_{13}$, due to collisions with the buffer gas, transit time and spin-exchange collisions. $\gamma_z$ is the relaxation rate of a population difference between Zeeman sublevels (labeled 2, 3, 4 in Fig.~\ref{fig2}) of a same hyperfine level towards the thermal equilibrium. In this calculation, we also take the response time of each device into consideration,
i.e., \SI{50}{\nano\second} for phase switch, \SI{2.5}{\micro\second} for polarization switch, \SI{0.5}{\micro\second} for
light chopper (AOM).

\subsection{\label{sec:level2}Numerical results}

\begin{figure}[h]
\centering
\includegraphics[width=1\linewidth]{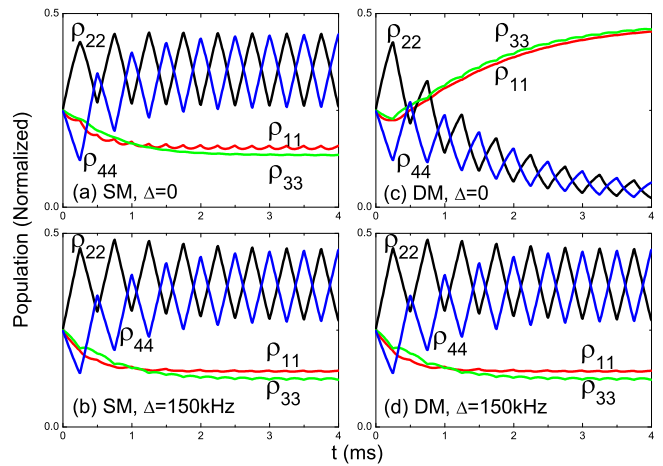}
\caption{(Color online) The evolution of ground-state population on ($\Delta=0$) and off ($\Delta=150$ kHz) Raman resonance in SM and DM cases.
The values of the parameters are:
$\Omega=4.4\times 10^6$\,\si{\per\second},
$\Gamma=2.6\times 10^9$\,\si{\per\second},
$\gamma_2=100$\,\si{\per\second},
$\gamma_1=\gamma_z=0.1\gamma_2$,
$f_m=2$\,\si{\kilo\hertz},
the signal is summed during a time $t_{w1}=$\,\SI{10}{\micro\second}.
$z=L/2=25$\,\si{\milli\meter},
atomic density: $n_a=8.8\times 10^{10}$\,\si{\per\cubic\centi\metre}
(corresponding to the atomic vapor cell temperature \SI{32.5}{\degreeCelsius}).}
\label{fig3}
\end{figure}

The evolution of ground-state population computed in the center of the cell illuminated with SM and DM light is
presented in Fig.~\ref{fig3}. Both the on ($\Delta=0$) and off ($\Delta=150$ kHz) Raman resonance cases are plotted. The numerical calculation is performed for a polarization modulation frequency $f_m=2$ kHz.
In the SM case, the atomic population is alternately pumped into one of the two ``end states", the clock states are  depopulated.
While on Raman resonance in the DM case, due to the polarization modulation and the constructive build-up in the same dark state, the atomic population accumulates in the clock states in few modulation periods. Off Raman resonance, the population distribution in the ground states is similar to the SM case.
\begin{figure}[h]
\centering
\includegraphics[width=0.8\linewidth]{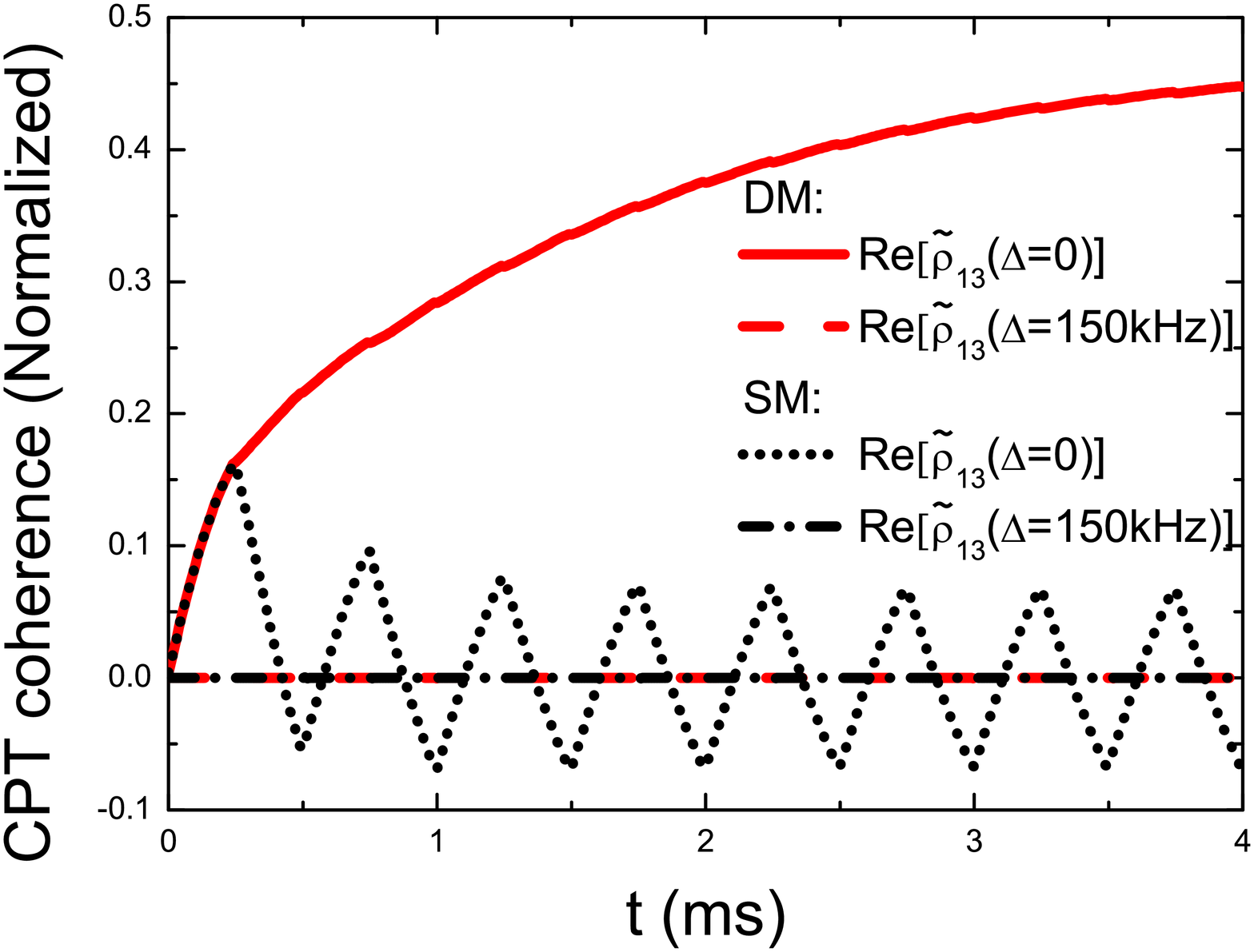}
\caption{(Color online). Evolution of the clock-state coherence $\rho_{13}$ on ($\Delta=0$) and off ($\Delta=150$ kHz) Raman resonance in SM and DM cases.
Same conditions as in Fig.3.}
\label{fig4}
\end{figure}

We can see the increase and the decrease of CPT dark states more clearly in Fig.~\ref{fig4},
which shows the real part of the clock-state coherence, also called CPT coherence, related to the optical absorption.
The CPT coherence is repeatedly built and destroyed in the SM case, while it continuously grows in the DM case.

\section{\label{sec:level1}Experimental results}
\subsection{\label{sec:level2}One pulse, one polarization}

\begin{figure}[htb]
\centering
\includegraphics[width=0.95\linewidth]{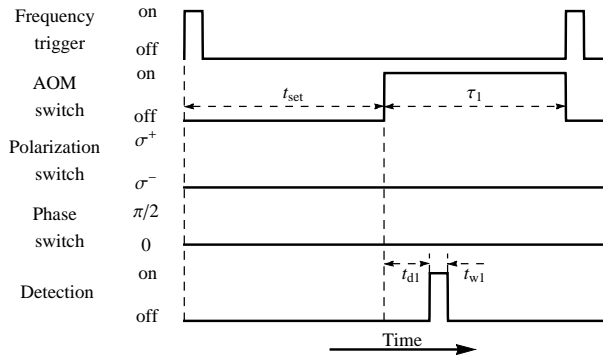}
\caption{Time sequence for $\sigma\sigma$ CPT.}
\label{fig5}
\end{figure}

In order to characterize two-frequency optical pumping effects, we have studied the dynamic response of the atomic vapor illuminated by a circularly polarized bichromatic beam. Thus we apply the time sequence presented in Fig.~\ref{fig5},
 in which a single circular $\sigma^-$ polarization and a constant Raman phase are used.
 Each measurement is composed of a pulse train. At each pulse, after triggering, we wait for
 a time $t_{set}\sim 100$\,\si{\milli\second} in the dark in order to allow full relaxation of the atomic
 system after the previous pulse. The lifetime of hyperfine coherences and population difference
 in the ground state is about $3$\,\si{\milli\second} in this vapor cell. Then the laser beam is switched on
 by means of the AOM during a time $\tau_1$. The optical power transmitted through the cell
 is recorded by the photodiode after an interrogation time or adjustable delay $t_{d1}$ and averaged
 during a time window $t_{w1}$.

\begin{figure}[t]
\centering
\subfigure[]{\includegraphics[width=0.9\linewidth]{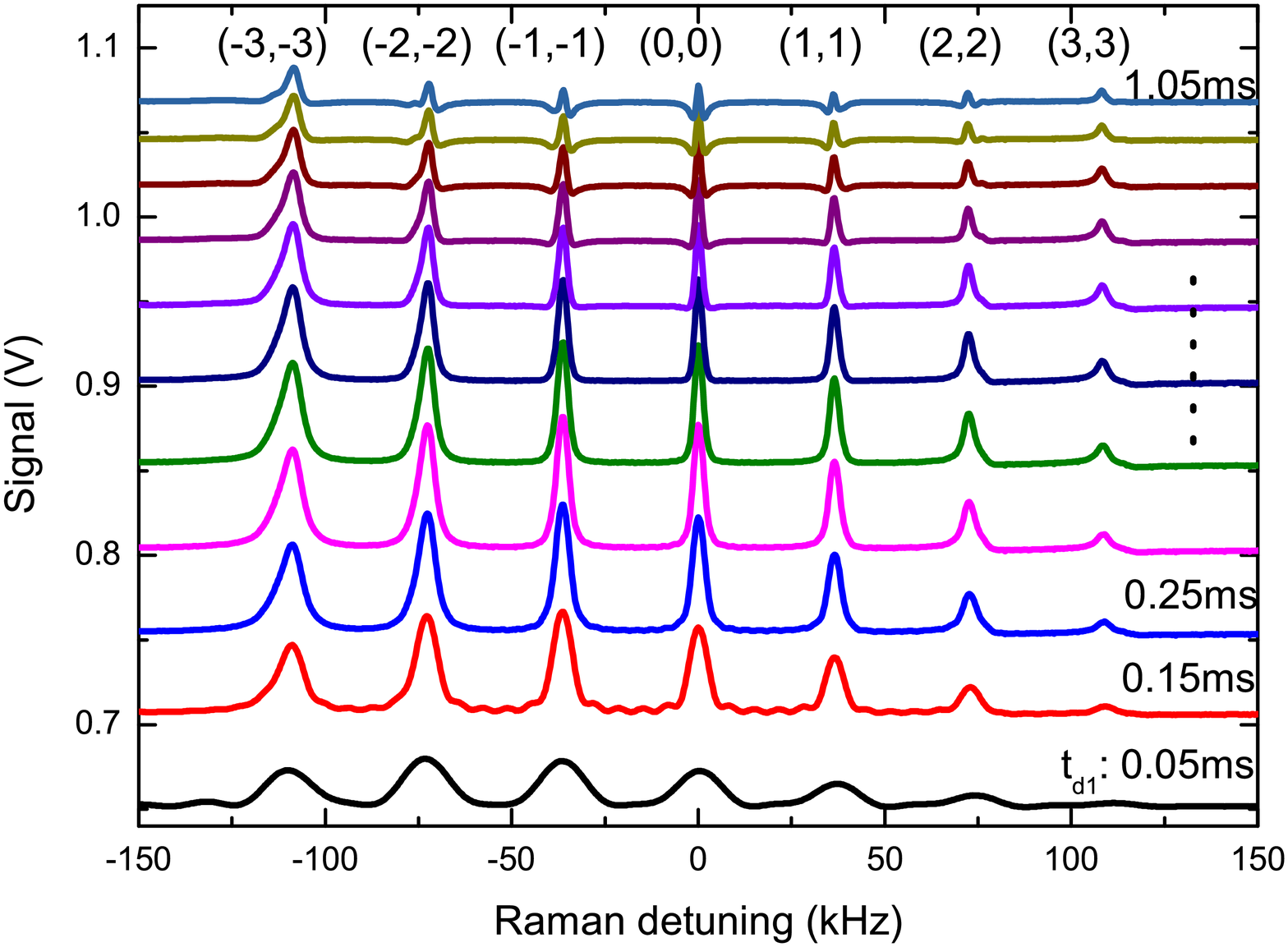}}
\label{fig6a}
\subfigure[]{\includegraphics[width=0.9\linewidth]{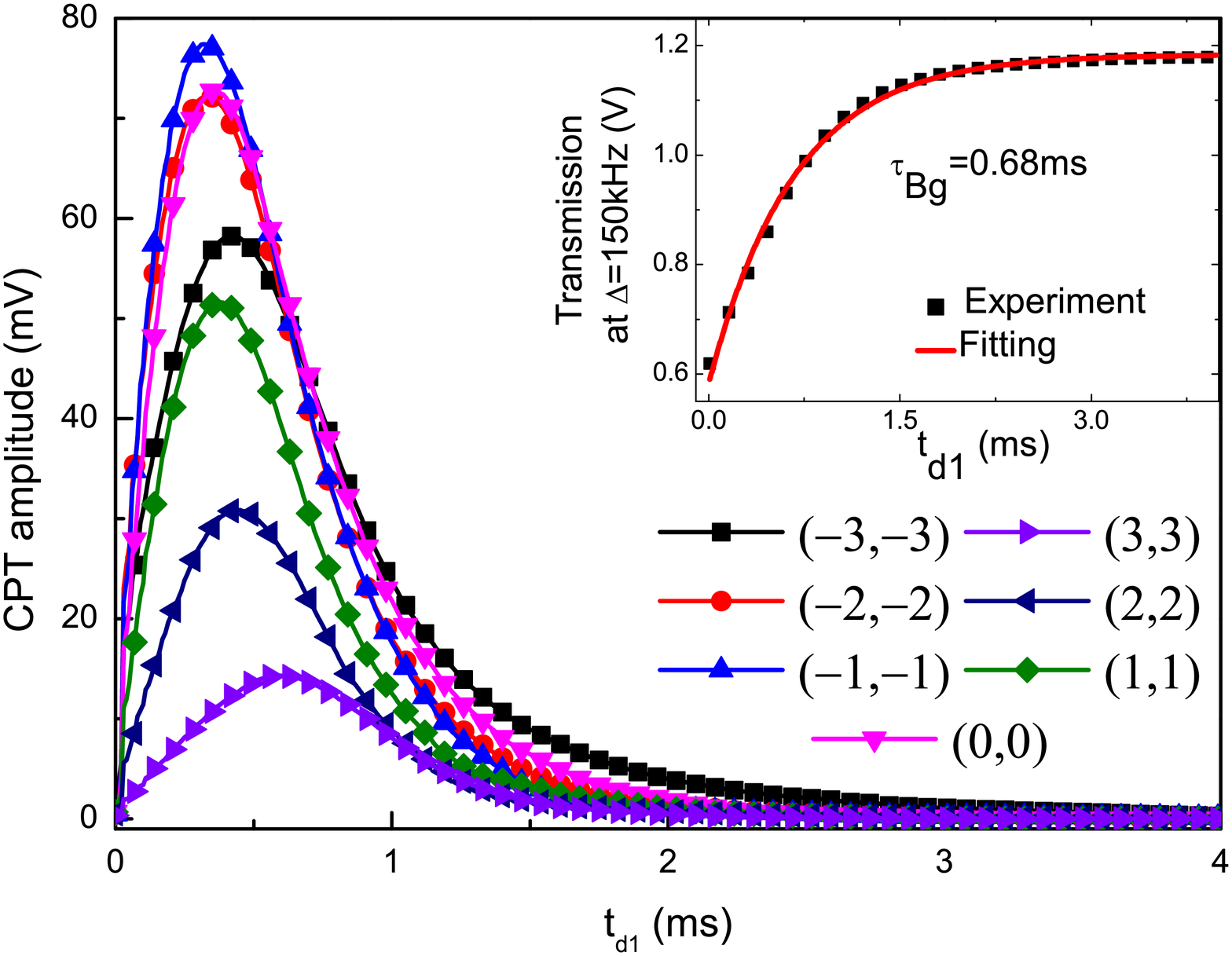}}
\label{fig6b}
\caption{(Color online)
(a) Zeeman CPT spectrum at various $t_{d1}$ values.
(b) $7$ CPT transition amplitudes as function of $t_{d1}$. Inset: background level as function of $t_{d1}$. $\gamma_{Bg}$ is an overall pumping rate. $\tau_{Bg}$ is the time constant given by fitting.
In all cases: $t_{set}=100$\,\si{\milli\second}, $t_{w1}=10$\,\si{\micro\second}, full laser power: $2.27$\,\si{\milli\watt}.
The laser polarization incident to cell is circularly polarized. }
\label{fig6}
\end{figure}

With $\sigma^-$ polarized light the only allowed optical transitions are transitions
with $m_F'- m_F=-1$, with $m_F'$ ($m_F$) the magnetic quantum number of the Zeeman sublevel of the excited (ground state) level, respectively. The CPT transitions in the ground state ($\ket{6^2 S_{1/2},F=3,m_F}\leftrightarrow\ket{6^2 S_{1/2},F=4,m_F}$, denoted ($m_F,m_F$) in the following, occur between sublevels of same $m_F$ when the two first sidebands frequency difference is equal to the ($m_F,m_F$) transition frequency. The Raman detuning $\Delta$ is the difference between the two first sidebands frequency gap minus the clock transition frequency. Fig.~\ref{fig6}a shows the CPT spectrum recorded by scanning the Raman detuning. The seven allowed CPT transitions are observed, from ($-3,-3$) to ($+3,+3$). As expected the peaks amplitude increase towards negative detuning, because atoms are optically pumped towards the end Zeeman sublevel, here $\ket{F=4,m_F=-4}$, and in part because the transition probabilities increase from right to left due to properties of dipole matrix elements \cite{Steck}. The higher $ \vert m_F \vert$ transitions are more broadened by the magnetic field inhomogeneity on the whole cell, as shown in Fig.~\ref{fig6}a, because they are more sensitive to the applied static Zeeman field. This broadening explains why the ($-3,-3$) peak is not the highest as expected. The spectrum is recorded for several delay time $t_{d1}$. Note that no offset is applied to separate the curves, it is the actual signal.

The evolution of the background level is plotted in the inset of Fig.~\ref{fig6}b. Here the Raman detuning is $\Delta=150$\,\si{\kilo\hertz}, which is out of resonance for the seven CPT resonances but still at resonance for one-photon resonances.
So that after several absorption-spontaneous emission cycles atoms are gradually pumped in the end
Zeeman sublevel where they cannot absorb light. At the steady-state the absorption reaches its minimum.
We fitted simply the time evolution of the background with an exponential law: $a-b\exp(-t/\tau_{Bg})$,
where $a,\, b$ are constants.
With this laser power the time constant is $\tau_{Bg}=0.69$\,\si{\milli\second} (pumping rate $\gamma_{Bg}=1/\tau_{Bg} = 1480$\,\si{\per\second}).
\begin{figure}[b!!]
\centering
\subfigure[]{\includegraphics[width=0.8\linewidth]{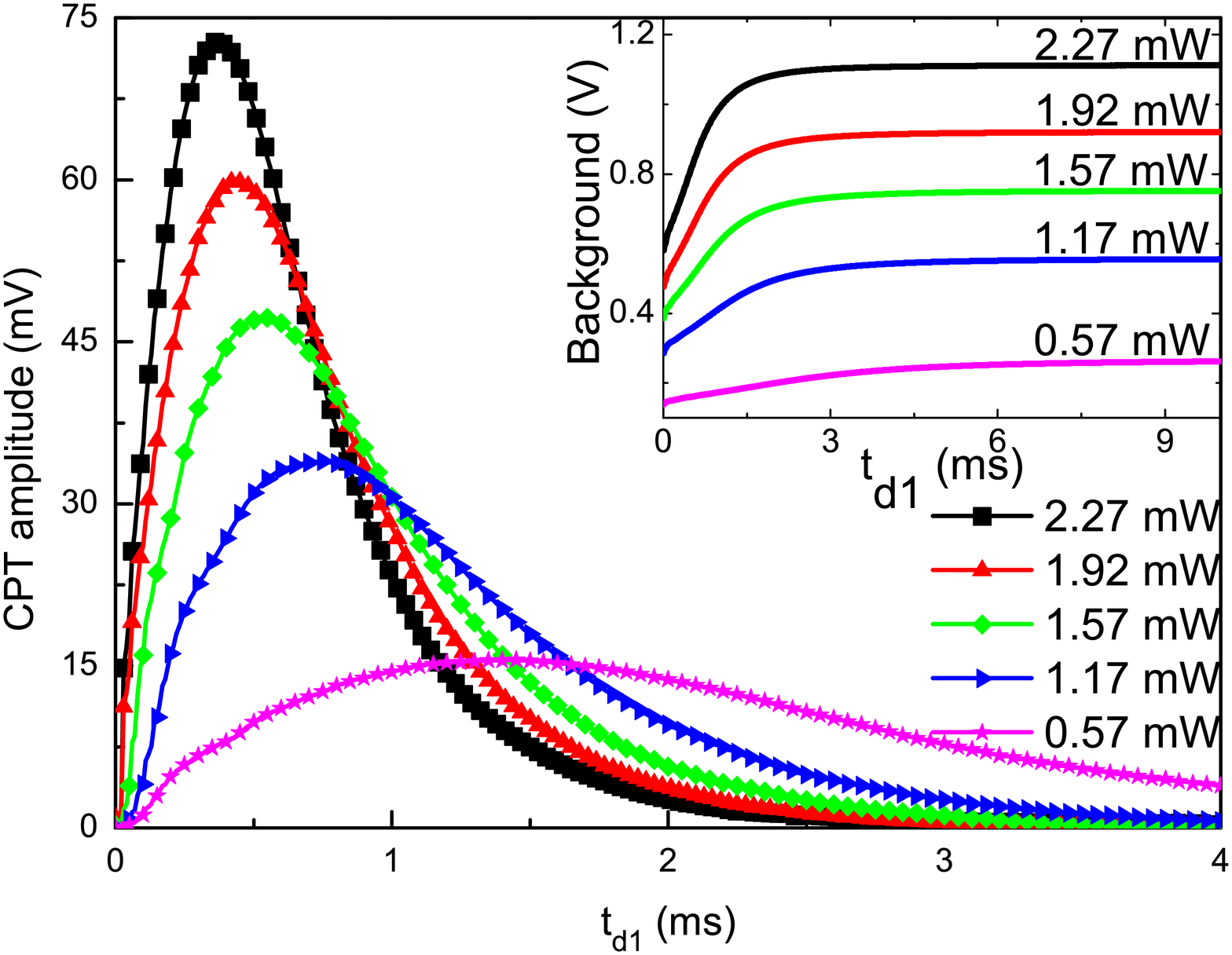}}
\subfigure[]{\includegraphics[width=0.9\linewidth]{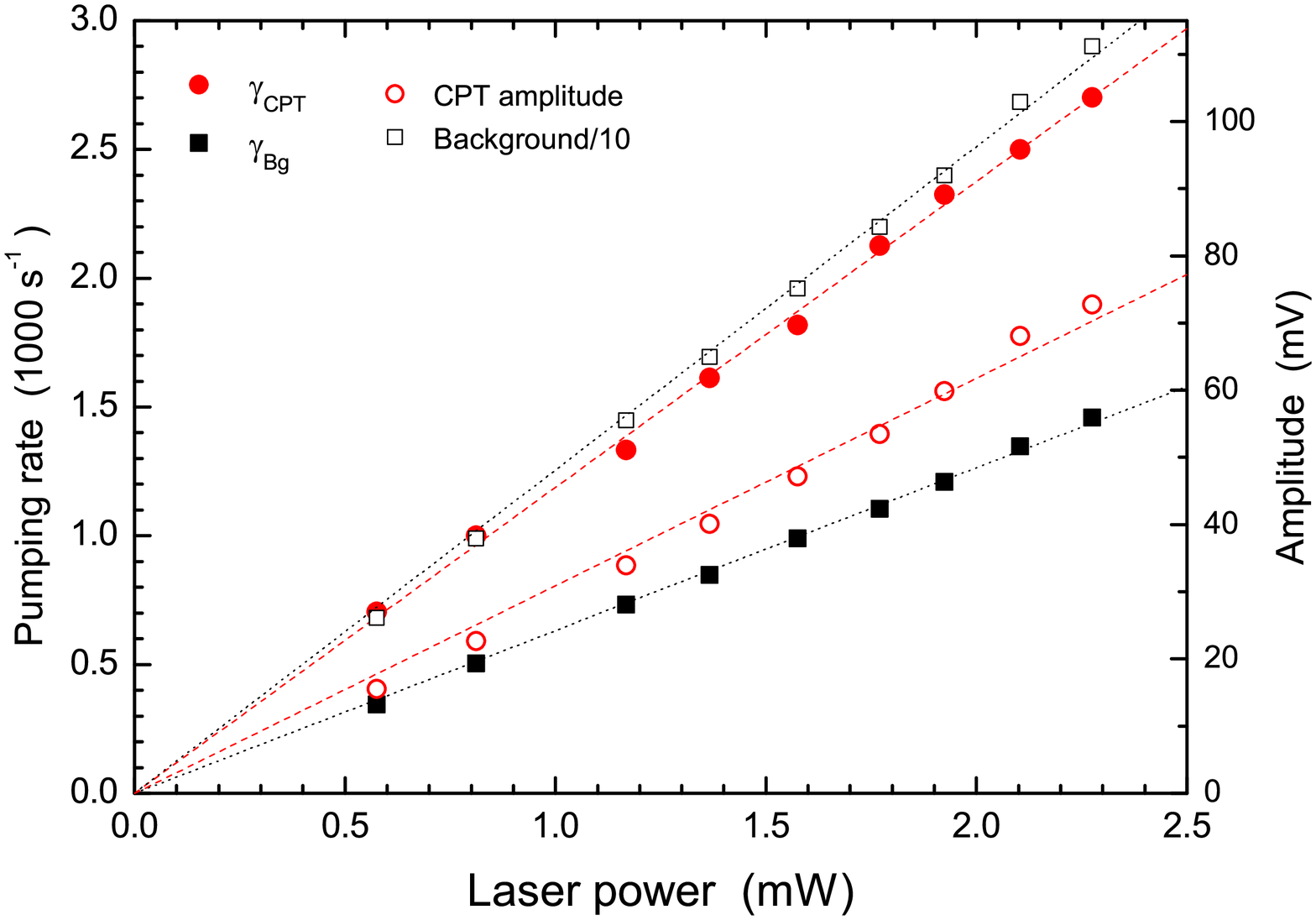}}
\caption{(Color online)The amplitude of the clock transition and background level (inset) as function of detection time $t_{d1}$ at various laser power.
(b) Left axis: CPT building  rate $\gamma_{CPT}$ and background pumping rate $\gamma_{Bg}$. Right axis: maximum CPT amplitude and steady-state background level (divided by $10$, $\Delta =150$\,\si{\kilo\hertz}) as function
of laser power. Dashed and dotted lines are linear fits.
In all cases, $t_{set}=100$\,\si{\milli\second}, $t_{w1}= 10$\,\si{\micro\second}.
}
\label{fig7}
\end{figure}

The dynamic behavior of the seven CPT transition amplitudes is also illustrated in Fig.~\ref{fig6}b. The CPT amplitude is the difference between the top of a CPT peak of Fig.\ref{fig6}a and the off-resonance background ($\Delta=150$\,\si{\kilo\hertz}).
All the seven CPT transition amplitudes follow the same behavior:
the first increase of the CPT amplitude is due to the faster pumping in the dark state than Zeeman and hyperfine pumping, where atoms undergo several absorption-emission cycles and circulate on several levels before reaching the end state.  Then when the effect of ``end-states" pumping overtakes the CPT pumping, the CPT amplitudes start to decrease. Finally when the ``end-states" pumping is near complete, the on-resonance absorption reaches a minimum value at the steady-state, which is close to the off-resonance minimum and leads to a negligible CPT amplitude.

The dynamic behavior of the clock transition and background at various laser power are shown in Fig.~\ref{fig7}. The maximum CPT amplitude increases with the laser power, and the time $t_{max}$ corresponding to this maximum becomes shorter (Fig.~\ref{fig7}a).
We define empirically $\gamma_{CPT}=1/t_{max}$, characteristic of the rate of building up the CPT state. This rate is approximately proportional to the laser power as shown in Fig.~\ref{fig7}b.
The background level (inset of Fig.~\ref{fig7}a) also increases with the laser power.
When the laser power is $2.27$\,\si{\milli\watt}, $t_{max} = 0.35$\,\si{\milli\second} while the time constant of the background level is $\tau_{Bg} = 0.69$\,\si{\milli\second}, this slower background evolution explains the time evolution of the CPT amplitude (Fig.~\ref{fig7}a).
The overall pumping rate  $\gamma_{Bg}$, the CPT building rate  $\gamma_{CPT}$, the amplitude maximum of the CPT transition and the steady-state value of the background level are all proportional to the laser power, as shown in Fig.~\ref{fig7}b.

\subsection{\label{sec:level2}One pulse, double-modulation}
 \subsubsection{Time sequence}

\begin{figure}[h]
\centering
\subfigure[]{\includegraphics[width=0.8\linewidth]{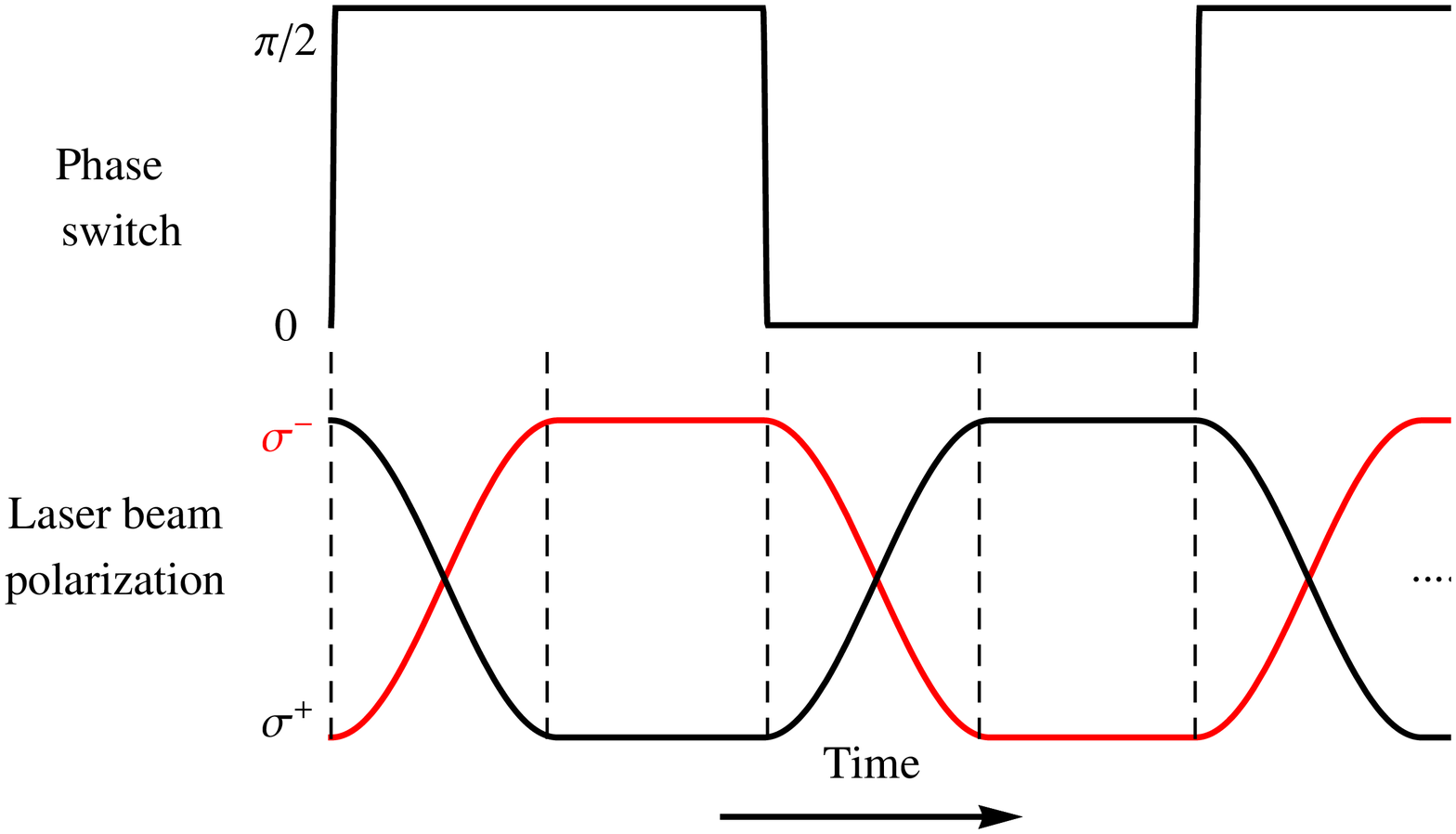}}
\subfigure[]{\includegraphics[width=0.95\linewidth]{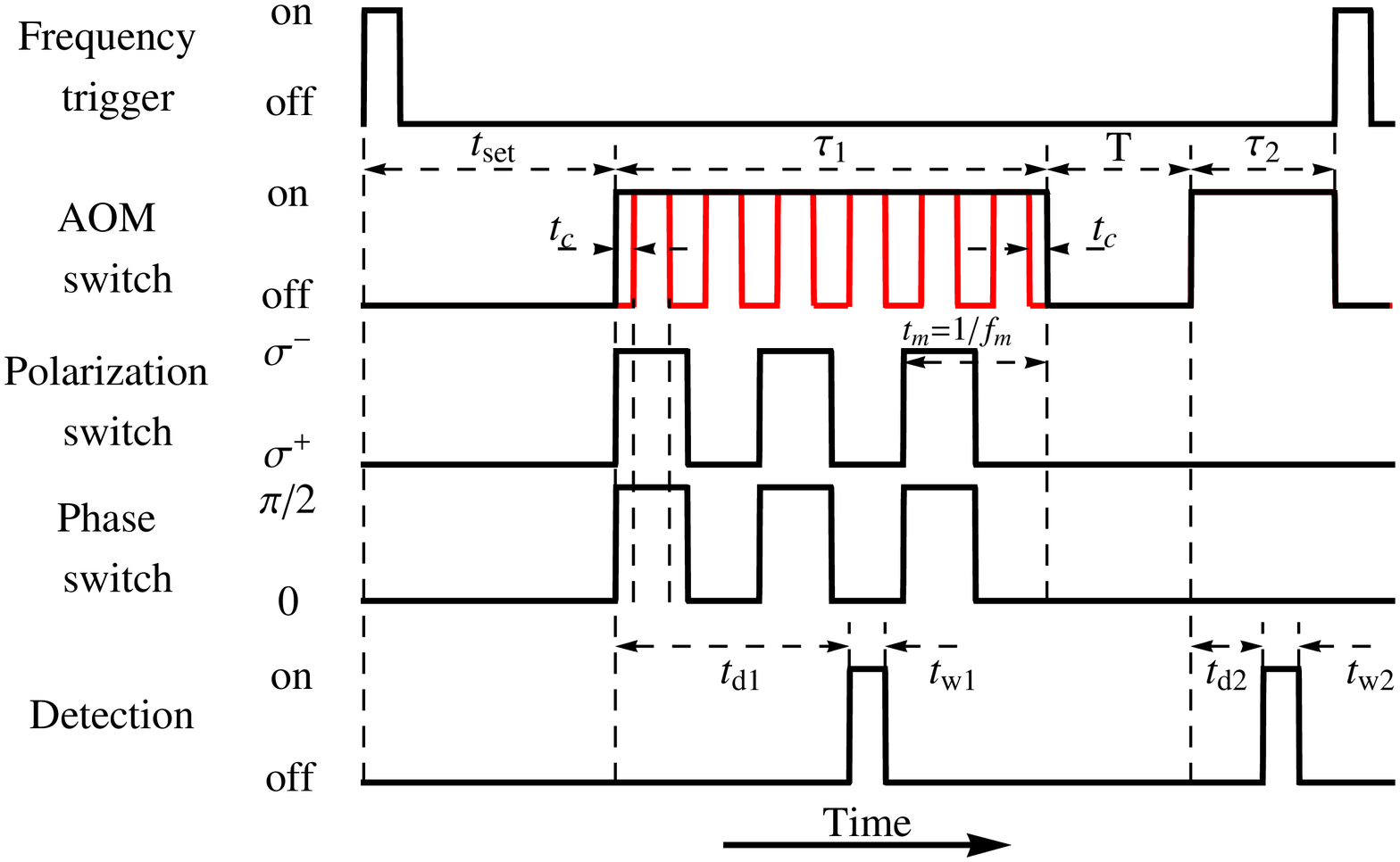}}
\caption{(Color online)
(a) Time response of phase and polarization switch (not to the scale).
(b)Time sequence for CPT experiments with (red) and without (black) clear up time.
Notice that there is no modulation during the probe pulse.}
\label{fig8}
\end{figure}

The polarization is now  modulated during the first optical pulse at the frequency $f_m$ (period $t_m$). During the rising or falling time of the polarization switch (about  $2.5$\,\si{\micro\second}) both counter-rotating circularly polarized beams coexist, while their Raman phase (phase switch time $<50$\,\si{\nano\second}) is the same as shown in Fig.~\ref{fig8}a. In this configuration the two build-up CPT states are in opposite phase and destroy each other \cite{Yun14}. The time sequence with clear up time $t_c$, used to investigate the double modulation scheme with one and two pulses, is shown in Fig.~\ref{fig8}b. When the polarization is switched, we turn the laser beam off during $t_c$ by means of the AOM. The new time sequence used to investigate the double modulation scheme with one and two pulses is shown in Fig.~\ref{fig8}b. For one-pulse investigation the signal is detected during the first pulse of length $\tau_1$. For two-pulse investigation the detection is performed during the second pulse of length $\tau_2$, the  polarization is not modulated during this second pulse  to avoid the effect induced by not ideal polarization switch. The remaining of the paper is devoted to the ($0, 0$) clock transition.

\subsubsection{Clock transition amplitude}

\begin{figure}[h]
\centering
%\resizebox{3.2in}{!}
\subfigure[]{\includegraphics[width=0.9\linewidth]{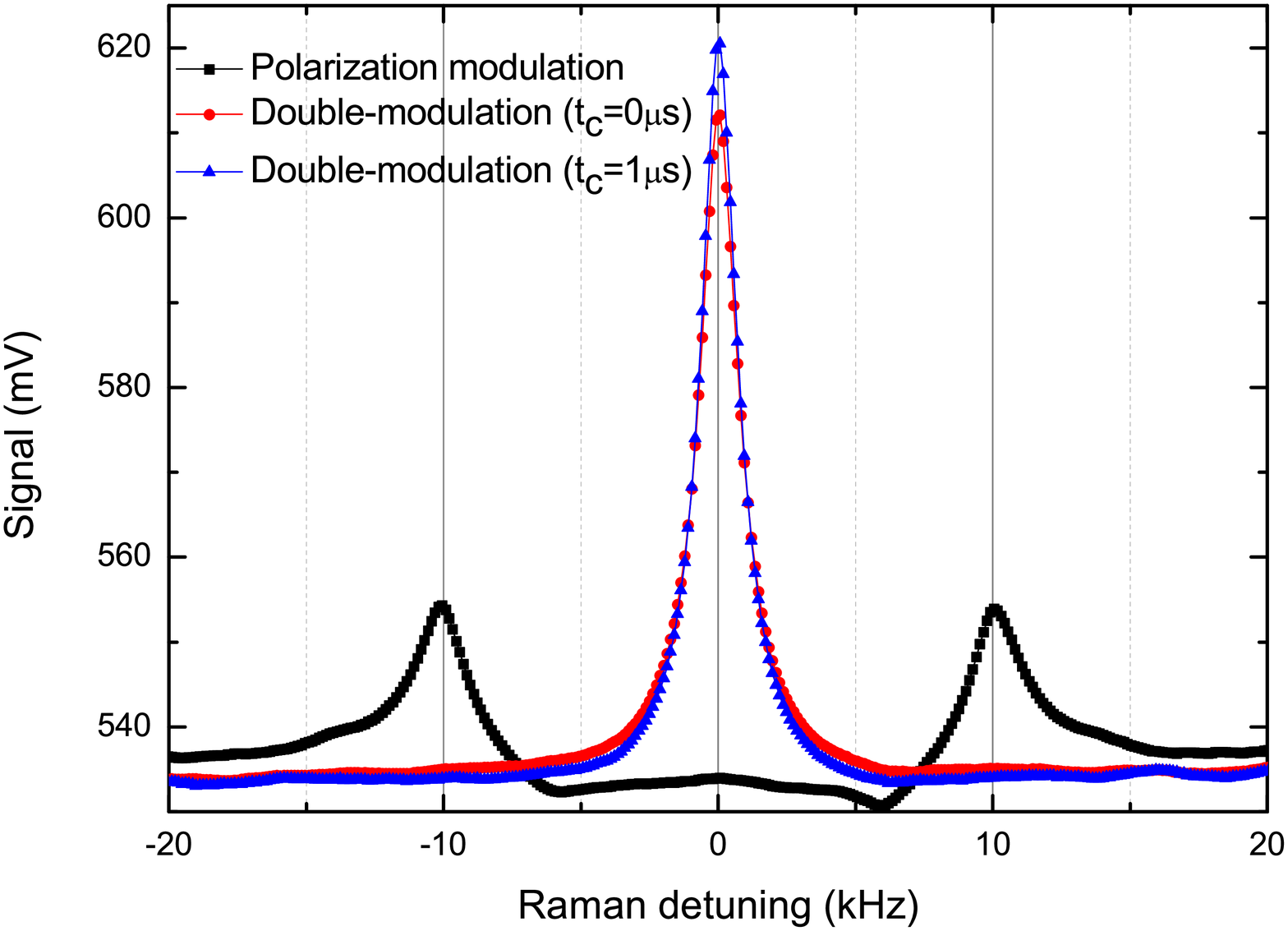}}
\subfigure[]{\includegraphics[width=0.9\linewidth]{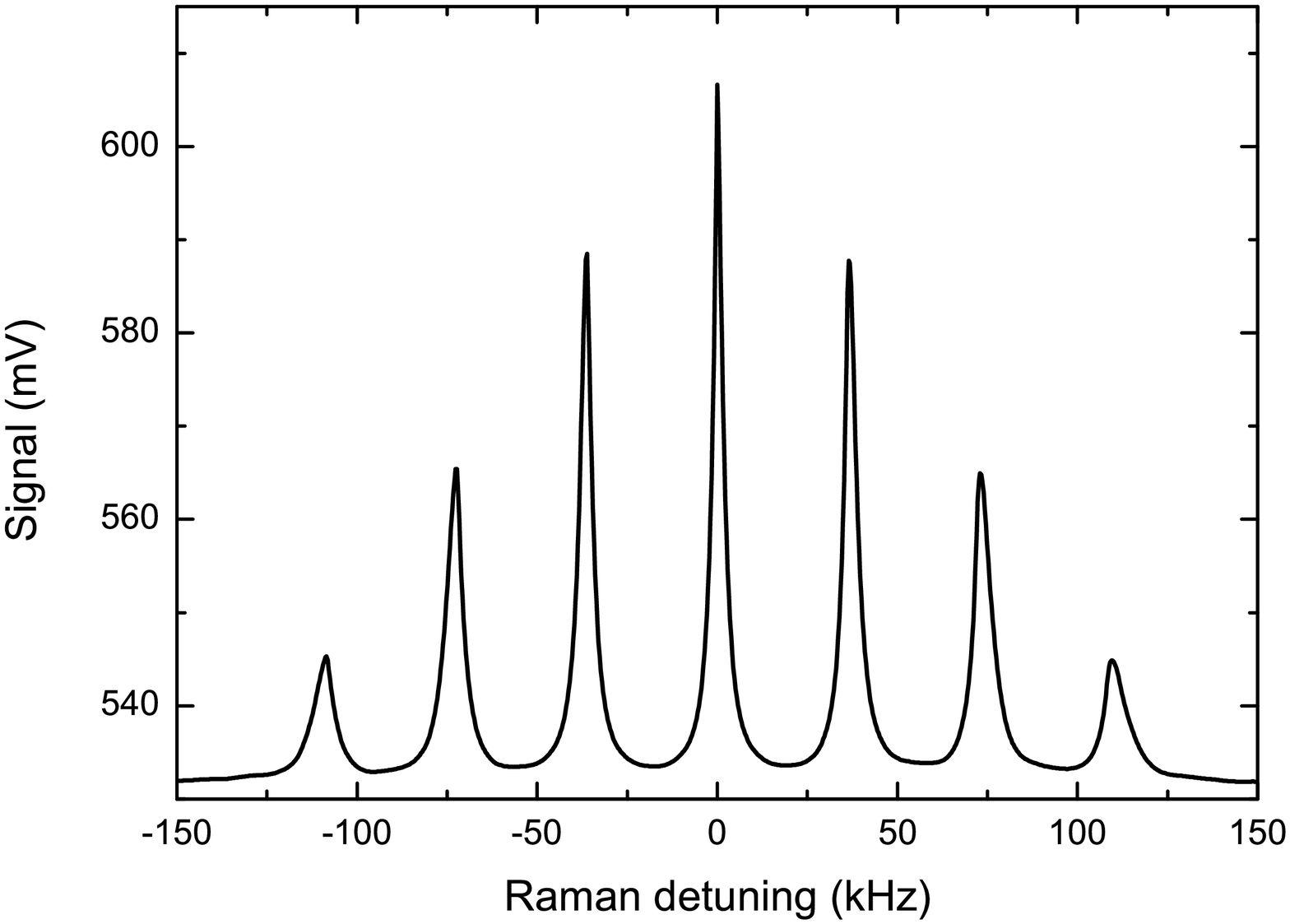}}
\caption{(Color online)
(a) CPT spectrum observed by polarization modulation and double-modulation scheme,
the latter includes two clear up times $t_c=0$\,\si{\micro\second} and $t_c=1$\,\si{\micro\second}, the detection window $t_{w1}=10$\,\si{\micro\second}.
(b) Zeeman CPT spectrum with double-modulation with  $t_c=0$\,\si{\micro\second} and $t_{w1}=100$\,\si{\micro\second}.
In both cases, $t_{set}=100$\,\si{\milli\second}, $t_{d1}=3.08$\,\si{\milli\second},
$f_m=10$\,\si{\kilo\hertz}, laser power: $2$\,\si{\milli\watt}.}
\label{fig9}
\end{figure}

The CPT spectrum with polarization modulation but without phase modulation is shown in Fig.~\ref{fig9}a. The extinction of the central peak clearly shows the mutual destruction of both dark states  \cite{Huang12, Yun14}. While the peak appears when the Raman phase is $\pi$ modulated synchronously with the polarization modulation (double modulation). This confirms the role of the Raman phase in the polarization modulation. With double modulation the spectrum is shown with and without the clear up time $t_c$. The signal is improved for $t_c=1$\,\si{\micro\second}, a contrast (CPT amplitude divided by background) of
$16.2\%$  and a linewidth of $1.5$\,\si{\kilo\hertz} is obtained. The full Zeeman spectrum is shown in Fig. \ref{fig9}b. Compared to Fig. \ref{fig6}a the efficiency of double modulation scheme is clearly visible, with population accumulation in the clock levels. The distortion of high $|m|$ peaks is explained by magnetic field inhomogeneity on the cell volume.

\begin{figure}[htb]
\centering
\includegraphics[width=1\linewidth]{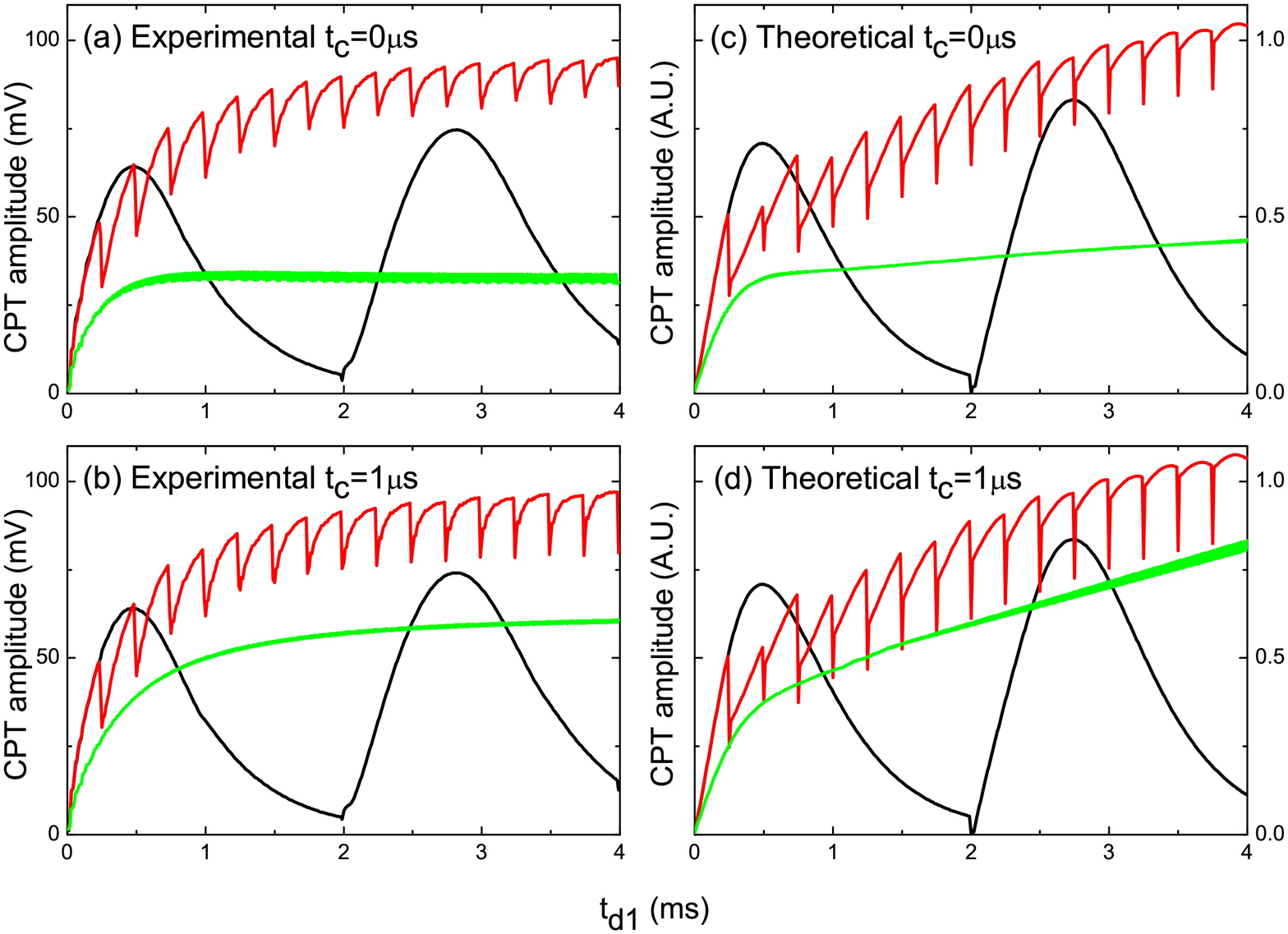}
\caption{(Color online)
Evolution of clock transition amplitude as function of the interaction time $t_{d1}$ obtained by experimental (a and b) and theoretical (c and d) methods.
In both case, we show $t_c=0$\,\si{\micro\second} (a and c) and $1$\,\si{\micro\second} (b and d).
In each cases, three value of $f_m$ are demonstrated: $0.25$\,\si{\kilo\hertz} (black, middle curve), $2$\,\si{\kilo\hertz} (red, top curve) and $50$\,\si{\kilo\hertz} (green, bottom curve).
Experimental parameters: $t_{set}=100$\,\si{\milli\second}, $t_{w1}=10$\,\si{\micro\second}, laser power: $2$\,\si{\milli\watt}.
Same calculation parameters as in Fig.3.
}
\label{fig10}
\end{figure}

The dynamic behavior of the CPT amplitude as function of the interaction time $t_{d1}$ for various modulation
frequencies $f_m$ with $t_c=0$\,\si{\micro\second} and $1$\,\si{\micro\second} is demonstrated in Fig.~\ref{fig10}, in which both the experimental (Fig.~\ref{fig10}a, b)
and theoretical (Fig.~\ref{fig10}c, d) results are presented.
At low modulation
frequency ($250$\,\si{\hertz}) the maximum signal increases at each half period.  At a laser power of $2$\,\si{\milli\watt} $t_{max}\approx 0.47$\,\si{\milli\second}, the overall characteristic pumping time is about $0.75$\,\si{\milli\second} (Fig.~\ref{fig7}).
The period is long enough to pump the atoms which are not in the dark state into the end Zeeman sublevel.
At each polarization reversal, these atoms are pumped into the opposite end Zeeman sublevel, and a part of them is trapped in the dark state, increasing the clock signal.
After a maximum of signal the systems tends to a new steady-state and the signal vanishes as in  Fig.~\ref{fig4}a. As $t_m>t_{max}$, a part of the cycle time is lost.
When $f_m=2$\,\si{\kilo\hertz}, $t_m \simeq t_{max}$, the accumulation to the dark state is the most efficient. At higher frequencies (\SIlist{50;100}{\kilo\hertz}) $t_m \ll t_{max}$, the pumping in the dark state is not efficient, since the atoms in the end sublevels have not the time to reach the clock sublevels. The effect of the clear up time $t_c$ is more visible at high modulation frequencies,
when the polarization switch time becomes comparable to $t_m$.
The theoretical signal is given by computing the laser intensities transmitted by the cell.
The numerical results presented in Fig.~\ref{fig10} (c and d)  are in very good agreement with the experimental results.

\begin{figure}[b]
\centering
\includegraphics[width=1\linewidth]{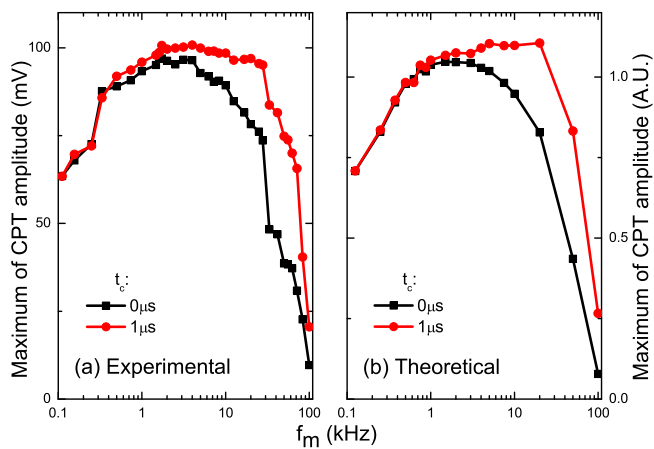}
\caption{(Color online)
Maximum of CPT amplitude as function of $f_m$ with $t_c=0$\,\si{\micro\second} and $1$\,\si{\micro\second} obtained by experimental (a) and theoretical (b) method.
Experimental parameters: $t_{set}=100$\,\si{\milli\second}, $\tau_1=4$\,\si{\milli\second}, $t_{w1}=10$\,\si{\micro\second}, laser power: $2$\,\si{\milli\watt}.
Same calculation parameters as in Fig.3.
}
\label{fig11}
\end{figure}

The experimental and theoretical maximum clock signals are presented in Fig.~\ref{fig11} as a function of $f_m$ for two values of the clear up time, $t_c=0$\,\si{\micro\second} and $1$\,\si{\micro\second}.
The maximum experimental CPT amplitude is increased with clear up time
$t_c=1$\,\si{\micro\second}. This advantage becomes obvious at high $f_m$ values.
However, further increase of $t_c$ makes this advantage less pronounced (not presented here). This is mainly caused by the shorter laser interaction time reducing the duty cycle.
Three regions of modulation frequency can be identified in this figure.
In the frequency range \SIrange{0.4}{30}{\kilo\hertz}, we call it the modulation frequency bandwidth, a high clock signal is observed. At lower values, $f_m<0.4$\,\si{\kilo\hertz}, the slow feeding of the dark state by polarization switch needs more cycles to reach the  maximum CPT amplitude steady-state. If the frequency is too low, there is no gain on the signal at each cycle. At high values, $f_m>30$\,\si{\kilo\hertz}, two effects reduce the signal amplitude. First, during the switch time of polarization modulation, without clear up time both counter-rotating circularly polarized beams coexist with the same Raman phase. This laser configuration induces a destructive interference for the two CPT states built by the two counter-rotating polarizations. The ratio of switch time to modulation period $t_m$ increases as $f_m$, thus the destructive effect of CPT states becomes more noticeable. With clear up time, the duty cycle is reduced.
Secondly, as explained above the high speed polarization switch is too fast compared to the atomic population pumping  time.
In such a case, the higher the $f_m$ value leads to a lower atomic population accumulation into the clock states. The theoretical results presented in Fig.~\ref{fig11}(b) match again the experimental results very well.

\begin{figure}[h]
\centering
\subfigure[]{\includegraphics[width=0.9\linewidth]{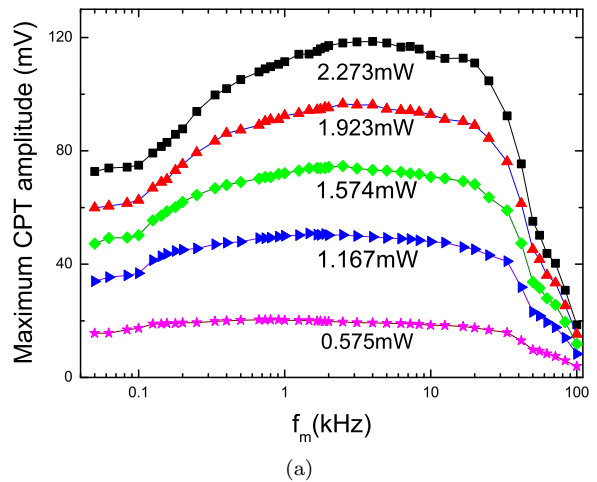}}
\label{fig9a}
\subfigure[]{\includegraphics[width=0.9\linewidth]{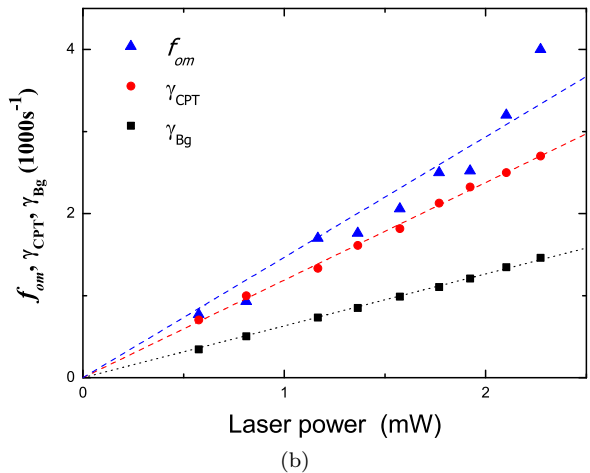}}
\label{fig9b}
\caption{(Color online)
(a) Maximum amplitude of the clock transition as function of $f_m$ for various laser powers.
(b)$\gamma_{Bg}$, $\gamma_{CPT}$ and $f_{om}$ as function of laser power. Dashed and dotted lines are linear fits.
In all cases, $t_{set}=100$\,\si{\milli\second}, $t_{w1}=10$\,\si{\micro\second}.
In the DM case $t_c=1$\si{\micro\second}.
}
\label{fig12}
\end{figure}

We have studied the maximum clock transition amplitude as a function of the modulation frequency for various laser powers as displayed in Fig.~\ref{fig12}a. Around the optimal value, the maximum amplitude exhibits a low sensitivity to the modulation frequency. The optimal modulation frequency $f_{om}$ versus the laser power is plotted in Fig.~\ref{fig12}b. It is roughly proportional to
the laser power as illustrated in Fig.~\ref{fig12}b with the following
empirical rule:  $\gamma_{Bg}<\gamma_{CPT}\leq f_{om}<min\{\Omega,1/2t_{EOM}\}$,
here $\Omega$ is the averaged Rabi frequency of our bichromatic laser.

\subsection{\label{sec:level2}Two-pulse, double-modulation. Ramsey spectrum}

\begin{figure}[htb]
\centering
\includegraphics[width=1\linewidth]{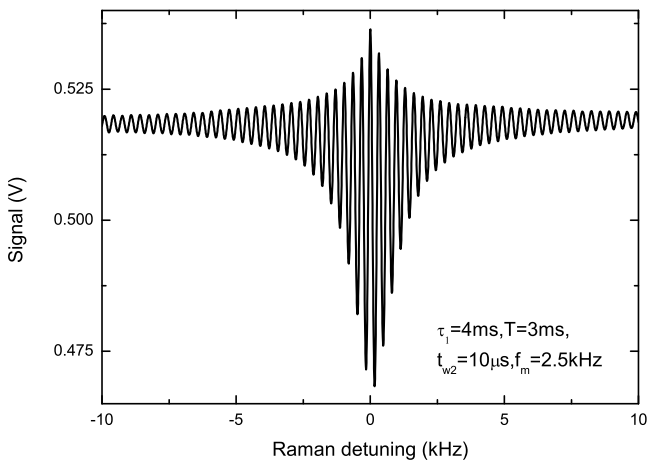}
\caption{Ramsey fringes. Parameters:
$t_{set}=100$\,\si{\milli\second}, $\tau_1=4$\,\si{\milli\second},
$t_{w2}=10$\,\si{\micro\second}, $t_{d2}=t_c=0$\,\si{\micro\second},
$f_m=2.5$\,\si{\kilo\hertz}, laser power: $2$\,\si{\milli\watt}.
Central fringe contrast: $13.5\,\%$, linewidth: $160$\,\si{\hertz}.
}
\label{fig13}
\end{figure}

Applying the time sequence shown in Fig.~\ref{fig8} with the detection during the second pulse, we get the two-pulse
Ramsey-CPT fringes shown in Fig.~\ref{fig13}. A signal with high contrast of  $13.5\%$ and linewidth
of $160$\,\si{\hertz} is obtained, promising for implementing a high performance CPT clock.
The asymmetry of the Ramsey fringes is due to an unbalance of the driving field amplitudes and an one-photon optical detuning.
As a change of these parameters could impact the mid-long term stability of the clock, they should be controlled in a working clock.
The effect of modulation frequency $f_m$ on the amplitude of the central Ramsey fringe and the $f_{om}$ dependence on the laser power in two-pulse Ramsey-CPT fringes are similar to the one-pulse case.

%\subsubsection{Ramsey fringe amplitude and time sequence parameters}

\begin{figure}[h]
\centering

\subfigure[]{\includegraphics[width=0.9\linewidth]{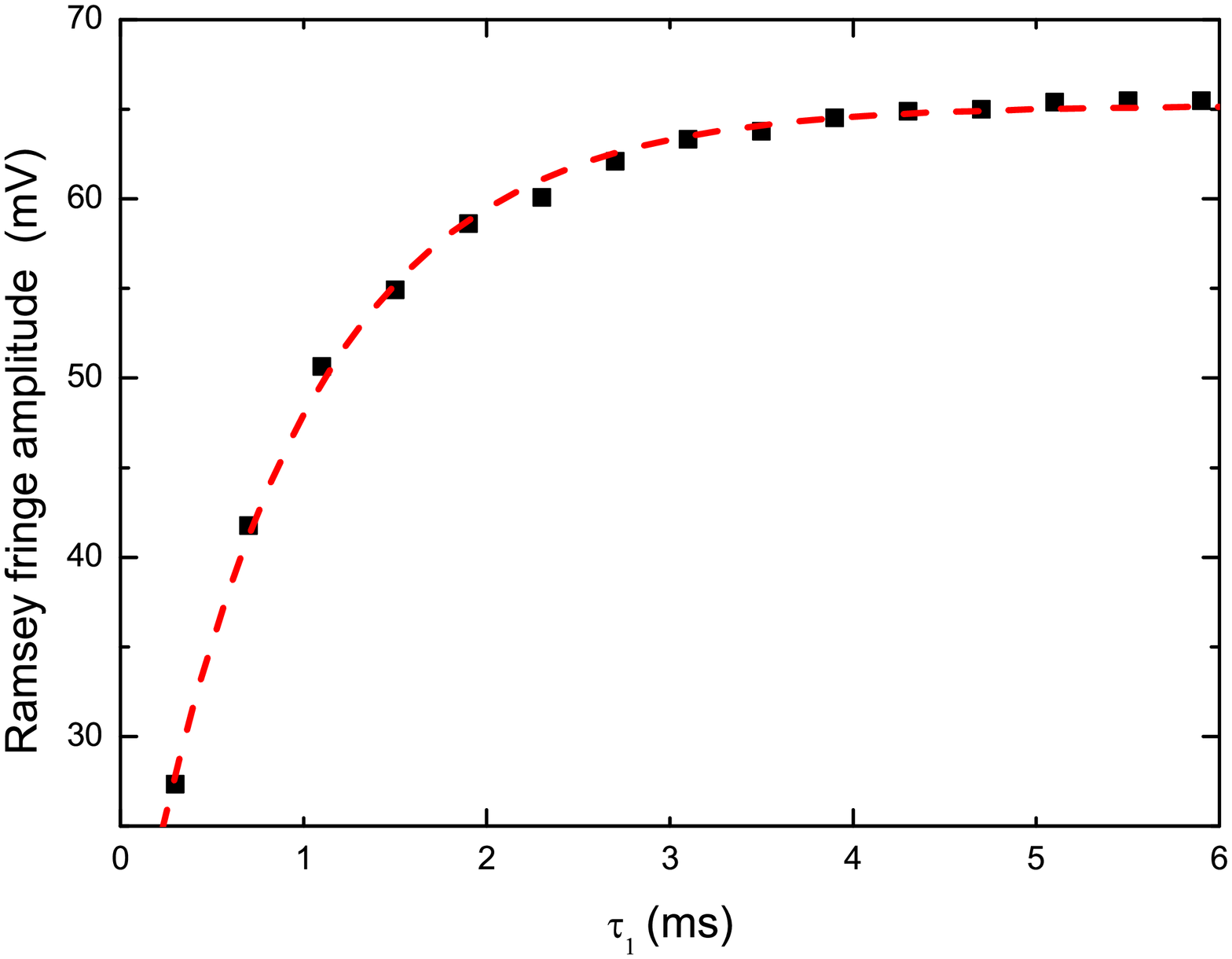}}
\label{fig14a}
\subfigure[]{\includegraphics[width=0.9\linewidth]{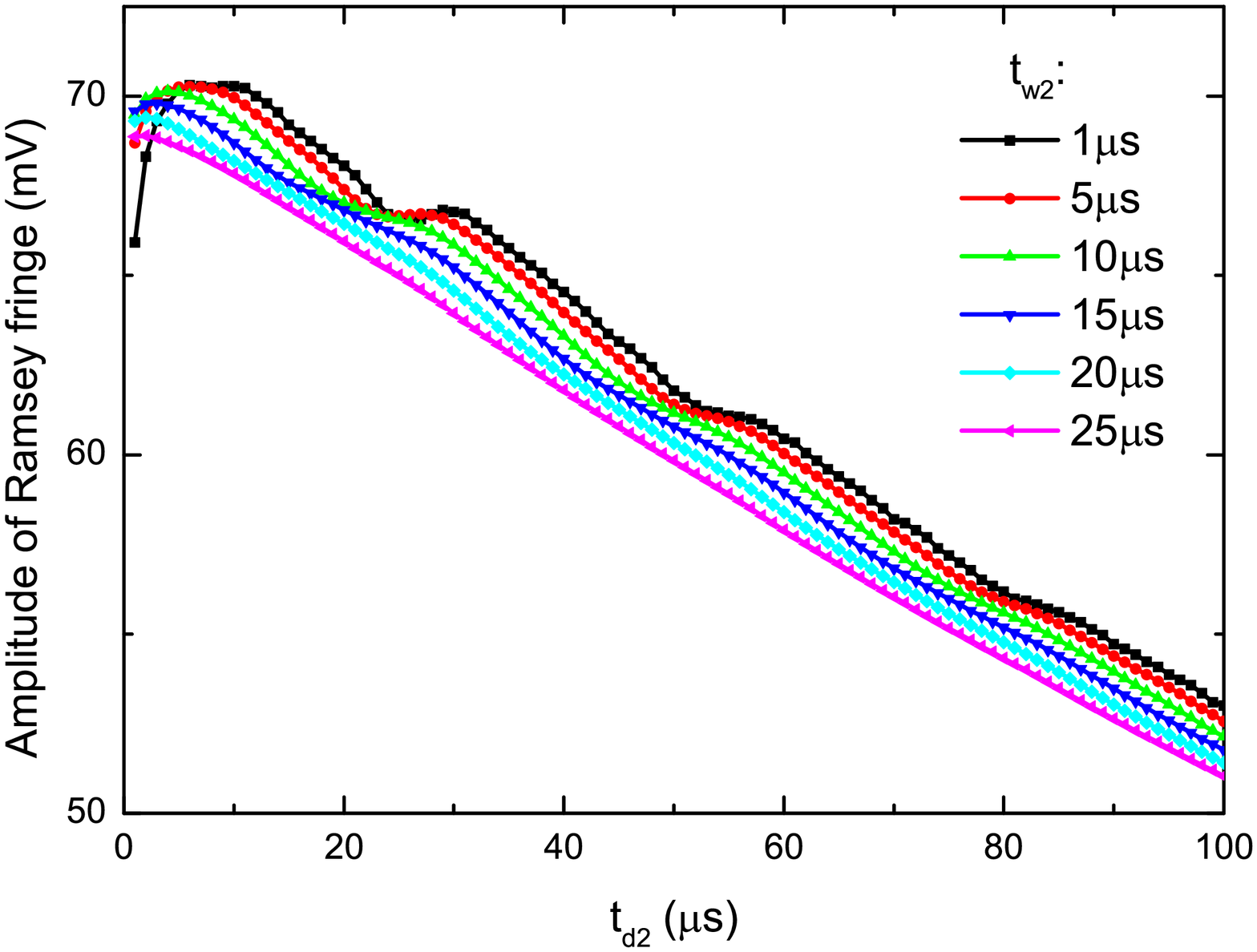}}
\label{fig14b}
\caption{(Color online)
Amplitude of Ramsey central fringe as function of
$\tau_1$  with $t_{d2}=0$\,\si{\micro\second} (a) and as function of
$t_{d2}$ with $\tau_1=4$\,\si{\milli\second} and various $t_{w2}$ (b).
The dashed line in (a) is a fit by an exponential law, the squares are the experimental data.
In both cases, $t_{set}=100$\,\si{\milli\second},
$T=3$\,\si{\milli\second}, $t_c=0$\,\si{\micro\second},
$f_m=2.5$\,\si{\kilo\hertz}, laser power: $2$\,\si{\milli\watt}.
}
\label{fig14}
\end{figure}

The effect of the first pulse length is presented in Fig.~\ref{fig14}a for a modulation frequency
$f_m=2.5$\,\si{\kilo\hertz}. The steady state is reached after more than ten modulation periods,
$\tau_1=\SI{4}{\milli\second} = 10\, t_m$. This value of  $\tau_1$ is similar to the one measured with crossed linear polarizations (lin-perp-lin configuration) \cite{Kozlova12}, the DM pumping is thus as efficient as the lin-perp-lin optical pumping configuration. Experimental points are fitted by an exponential law of time constant $0.91$\,\si{\milli\second}, the process is not much slower than the background pumping ($0.75$\,\si{\milli\second}).

Fig.~\ref{fig14}b shows the impact of the detection delay time on the Ramsey signal amplitude during the second pulse. The signal is maximized by a short delay $t_{d2}$ because for longer times the atomic system evolves towards its steady state, Ramsey fringes vanish, we come back to the one pulse case \cite{Zanon05, Zanon05b}. The diminution of Ramsey fringes over time explains also the decrease of the signal with the detection length $t_{w2}$.
For short detection windows $t_{w2}$, an oscillation appears on the probe signal with a period of about $28$\,\si{\micro\second} and decays quickly with a time constant of about $50$\,\si{\micro\second}. This is the transient response of the detuned adjacent CPT
transitions, ($-1,-1$) and ($1,1$), which are  $36$\,\si{\kilo\hertz} apart and oscillate at the detuning frequency (period $28$ \si{\micro\second}) \cite{Park04}.
The oscillation is no longer visible after a longer $t_{d2}$ value or averaged window time $t_{w2}$. Since this oscillation frequency is directly linked to the applied static magnetic field, it could be used in clock application for monitoring and locking the magnetic field value.

\begin{figure}[h]
\centering
\includegraphics[width=0.9\linewidth]{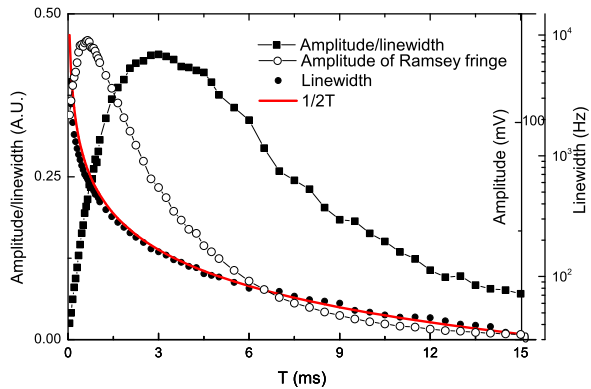}
\caption{(Color online)
The factor of merit, amplitude and linewidth of Ramsey central fringe as function of T.
Parameters: $t_{set}=100$\,\si{\milli\second},
$\tau_1=4$\,\si{\milli\second},
$t_{d2}=t_c=0$\,\si{\micro\second},
$t_{w2}=10$\,\si{\micro\second},
$f_m=2.5$\,\si{\kilo\hertz},
laser power: $2$\,\si{\milli\watt}.
}
\label{fig15}
\end{figure}

As expected, the linewidth of the Ramsey fringe scales as $1/2T$ (Fig.~\ref{fig15}),
with  $T$ the Ramsey time between the two pulses. During this time populations and
hyperfine coherence relax, the signal then vanishes for long $T$ values.
At short $T$ values the signal also decreases because a minimum $T$ value is needed to
establish the Ramsey fringes. When limited by a white frequency noise the clock frequency
stability scales as the linewidth-to-signal ratio. The inverse signal-to-linewidth ratio is often taken as a figure of merit.
As shown in Fig.~\ref{fig15}, it is maximized at $T=3.3$\,\si{\milli\second}.
This value corresponds to the hyperfine coherence lifetime in the cell, as expected \cite{Guerandel07}.

\section{\label{sec:level1}Conclusion}

\par We have shown that the CPT transition amplitude obtained with a single circularly polarized bichromatic field increases with time until a maximum, before vanishing because of longer response time of the Zeeman pumping. The backgound level meets the resonance top at the steady-state.
Zeeman pumping can be circumvented by polarization modulation. However, due to the relative phase
(see the Clebsch-Gordan coefficients in ref.~\onlinecite{Steck}) of the electric dipole moments involved in the transitions, an additional synchronous Raman phase modulation is needed to generate a constructive and highly-contrasted CPT resonance. With this double modulation technique, a non-zero signal is obtained at the steady state, except for low or high modulation frequencies.
The optimal modulation frequency $f_{om}$ is found approximately proportional to the optical pumping rate and satisfying this relation: $\gamma_{Bg}<\gamma_{CPT}\leq f_{om}<min\{\Omega,1/2t_{EOM}\}$.
The typical range of $f_{om}$ at a few \si{\kilo\hertz} releases the demand on the polarization switch time. Experimental results are in good agreement with a simple seven-level theoretical model.
The double modulation technique can be combined  with a pulsed (Ramsey) interrogation method. In this case, the polarization is not modulated during the detection pulse. The high contrast and narrow linewidth observed in Ramsey-CPT interrogation show the potential of double modulation to build a high performance CPT clock. Meanwhile, we can also maintain the compactness and robustness of CPT clock, e.g., by directly modulating a high modulation bandwidth DFB laser diode to remove the EOPM, and utilizing a liquid crystal polarization switch \cite{Yun15} to replace the EOM. The optimized parameters presented here are a guide-line for such a clock. This constructive polarization modulation method is attractive in atomic clock, atomic magnetometer and high precision spectroscopy applications.

\section*{Acknowledgements}
\par
We gratefully acknowledge Thomas Zanon-Willette for a careful reading of the manuscript and constructive suggestions. We would also like to thank Rodolphe Boudot, David Holleville, Fran\c{c}ois Tricot, Sinda Mejri, Natascia Castagna, Jean-Marie Danet, and Luca Lorini for helpful discussions.
We are grateful to Jos\'e Pinto Fernandes, Michel Lours and Electronic service at SYRTE for technical assistance and the realization of various electronic devices, Pierre Bonnay and Annie G\'erard for manufacturing Cs cells. P.Y. thanks Alexander Franzen for his ComponentLibrary.
\par This work is supported in part by ANR and DGA (ISIMAC project ANR-11-ASTR-0004). This work has been funded by the EMRP program (IND55 Mclocks). The EMRP is jointly funded by the EMRP participating countries within EURAMET and the European Union.

\appendix
\section{MAXWELL-LIOUVILLE EQUATION SETS}
%\numberwithin{equation}{section}
%\renewcommand{\theequation}{A.\arabic{equation}}

We consider one dimension problem, assume the light propagating along the z-axis, the atomic vapor cell is between the position  $z=0$ and $z=L$, with a cell length $L=50$ \si{\milli\meter}. The equation for the electric-dipole interaction of light field with the atomic ensemble is \cite{Scully}:
\begin{eqnarray}
&\frac{\partial^2 \vect{E}}{\partial z^2}-\frac{1}{c^2}\frac{\partial^2 \vect{E}}{\partial t^2}=\mu_0 \frac{\partial^2 \vect{P}}{\partial t^2}.
\label{vectP}
\end{eqnarray}
where $\vect{P}$ is the atomic ensemble polarization, $\vect{P}=n_a Tr(\rho d)$ ($n_a$ is the atomic density). From
 Eq.~(\ref{Hamiltonian},\ref{tilderho},\ref{V},\ref{Omega},\ref{vectP}), we find:

\begin{widetext}
%\begin{eqnarray}
\begin{align}
&\dot{\rho}_{11}=\Omega_{15}Im(e^{-i(\varphi+\phi+\pi/2)}\tilde\rho_{15})+\Omega_{17}Im(e^{-i(\varphi+\phi)}\tilde\rho_{17})
+(\rho_{55}+\rho_{66}+\rho_{77})\Gamma/3+(\rho_{22}+\rho_{33}+\rho_{44}-3\rho_{11})\gamma_1, \nonumber\\
&\dot{\rho}_{22}=-\Omega_{26}Im(e^{-i(\varphi-\phi)}\tilde\rho_{26})+(\rho_{55}+\rho_{66})\Gamma/3
+(\rho_{11}-\rho_{22})\gamma_1+(\rho_{33}-\rho_{22})\gamma_z,   \nonumber\\
&\dot{\rho}_{33}=-\Omega_{35}Im(e^{-i(\varphi-\phi+\pi/2)}\tilde\rho_{35})-\Omega_{37}Im(e^{-i(\varphi-\phi)}\tilde\rho_{37})
+(\tilde\rho_{55}+\tilde\rho_{77})\Gamma/3+(\tilde\rho_{11}-\tilde\rho_{33})\gamma_1
+(\tilde\rho_{22}+\tilde\rho_{44}-2\tilde\rho_{33})\gamma_z, \nonumber\\
&\dot{\rho}_{44}=-\Omega_{46}Im(e^{-i(\varphi-\phi+\pi/2)}\tilde\rho_{46})+(\rho_{66}+\rho_{77})\Gamma/3
+(\rho_{11}-\rho_{44})\gamma_1+(\rho_{33}-\rho_{44})\gamma_z ,\nonumber\\
&\dot{\rho}_{55}=-\Omega_{15}Im(e^{-i(\varphi+\phi+\pi/2)}\tilde\rho_{15})
+\Omega_{35}Im(e^{-i(\varphi-\phi+\pi/2)}\tilde\rho_{35})-\rho_{55}\Gamma, \nonumber\\
&\dot{\rho}_{66}= \Omega_{26}Im(e^{-i(\varphi-\phi)}\tilde\rho_{26})
+\Omega_{46}Im(e^{-i(\varphi-\phi+\pi/2)}\tilde\rho_{46})-\rho_{66}\Gamma, \nonumber\\
&\dot{\rho}_{77}= -\Omega_{17}Im(e^{-i(\varphi+\phi)}\tilde\rho_{17})
+\Omega_{37}Im(e^{-i(\varphi-\phi)}\tilde\rho_{37})-\rho_{77}\Gamma, \nonumber\\
&\dot{\tilde\rho}_{13}=
[e^{i(\varphi+\phi+\pi/2)}\Omega_{15}\tilde\rho_{35}^* +e^{i(\varphi+\phi)}\Omega_{17}\tilde\rho_{37}^*
+e^{-i(\varphi-\phi+\pi/2)}\Omega_{35}\tilde\rho_{15}+e^{-i(\varphi-\phi)}\Omega_{37}\tilde\rho_{17}]i/2
-(i\Delta+\gamma_2)\tilde\rho_{13}, \nonumber\\
&\dot{\tilde\rho}_{15}= [e^{i(\varphi+\phi+\pi/2)}\Omega_{15}(\rho_{55}-\rho_{11})+e^{i(\varphi-\phi+\pi/2)}\Omega_{35}\tilde\rho_{13}]i/2-(i\Delta_1+\gamma_{15})\tilde\rho_{15}, \nonumber\\
&\dot{\tilde\rho}_{17}= [e^{i(\varphi+\phi)}\Omega_{17}(\rho_{77}-\rho_{11})+e^{i(\varphi-\phi)}\Omega_{37}\tilde\rho_{13}]i/2-(i\Delta_1+\gamma_{17})\tilde\rho_{17}, \nonumber\\
&\dot{\tilde\rho}_{26}= e^{i(\varphi-\phi)}\Omega_{26}(\rho_{22}-\rho_{66})i/2-(i\Delta_2+\gamma_{26})\tilde\rho_{26}, \nonumber\\
&\dot{\tilde\rho}_{35}= [e^{i(\varphi-\phi+\pi/2)}\Omega_{35}(\rho_{33}-\rho_{55})-e^{i(\varphi+\phi+\pi/2)}\Omega_{15}\tilde\rho_{13}^*]i/2-(i\Delta_2+\gamma_{35})\tilde\rho_{35}, \nonumber\\
&\dot{\tilde\rho}_{37}= [e^{i(\varphi-\phi)}\Omega_{37}(\rho_{33}-\rho_{77})-e^{i(\varphi+\phi)}\Omega_{17}\tilde\rho_{13}^*]i/2-(i\Delta_2+\gamma_{37})\tilde\rho_{37}, \nonumber\\
&\dot{\tilde\rho}_{46}= e^{i(\varphi-\phi+\pi/2)}\Omega_{46}(\rho_{44}-\rho_{66})i/2-(i\Delta_2+\gamma_{46})\tilde\rho_{46}, \nonumber\\
&\frac{\partial \Omega_{15}}{\partial z}+\frac{1}{c}\frac{\partial \Omega_{15}}{\partial t}=\alpha Im[e^{-i(\varphi+\phi+\pi/2)}\tilde\rho_{15}], \nonumber\\
&\frac{\partial \Omega_{35}}{\partial z}+\frac{1}{c}\frac{\partial \Omega_{35}}{\partial t}=\alpha Im[-e^{-i(\varphi-\phi+\pi/2)}\tilde\rho_{35}], \nonumber\\
&\frac{\partial \Omega_{46}}{\partial z}+\frac{1}{c}\frac{\partial \Omega_{46}}{\partial t}=\alpha Im[-e^{-i(\varphi-\phi+\pi/2)}\tilde\rho_{46}], \nonumber\\
&\frac{\partial \Omega_{17}}{\partial z}+\frac{1}{c}\frac{\partial \Omega_{17}}{\partial t}=\alpha Im[e^{-i(\varphi+\phi)}\tilde\rho_{17}], \nonumber\\
&\frac{\partial \Omega_{37}}{\partial z}+\frac{1}{c}\frac{\partial \Omega_{37}}{\partial t}=\alpha Im[-e^{-i(\varphi-\phi)}\tilde\rho_{37}], \nonumber\\
&\frac{\partial \Omega_{26}}{\partial z}+\frac{1}{c}\frac{\partial \Omega_{26}}{\partial t}=\alpha Im[-e^{-i(\varphi-\phi)}\tilde\rho_{26}].
\end{align}
%\end{eqnarray}
\label{Maxwell-Liouville-1}
\end{widetext}

Here $\alpha=\omega_0 n_a d^2/(\epsilon_0 \hbar c)$,
$\Delta_{1(2)}=\omega_{+(-)} - \omega_{51(53)}$ are the optical detunings,  $\omega_{ij}=\omega_i-\omega_j$. The Raman detuning is $\Delta=\Delta_1-\Delta_2$.
In our case, we have $\Delta_1=-\Delta_2=\Delta/2$.
To get Eq.~(\ref{Maxwell-Liouville-1}2), we have neglected all coherences both in the ground and excited states,
except optical coherences and the clock-state coherence.
The phase difference between both driving fields evolves along the z-axis like $(k_+ - k_-) z$,
where $k_+$ and $k_-$ are the wavenumbers of the fields  $\omega_+$ and  $\omega_-$. The atoms,
confined by the buffer gas, are pumped in the corresponding dark state,
the phase of which is specific to their position. As the atoms evolve on
short distances compared to the phase-difference period of \SI{33}{\milli\meter} during
times of a few ms considered here, we can ignore this phase variation in the computation, without loss of generality.\\

Since the Rabi frequency $\Omega_{46(26)}$ is generated by the same laser field as $\Omega_{35(37)}$, we need to merge them.
With the adiabatic approximation:
$\dot{\tilde\rho}_{15}=\dot{\tilde\rho}_{35}=\dot{\tilde\rho}_{26}=\dot{\tilde\rho}_{46}=\dot{\tilde\rho}_{17}=\dot{\tilde\rho}_{37}=0$,
 and $\Delta\ll \Gamma$,$\gamma_{ij}=\Gamma/2$, we obtain:

\begin{widetext}
\begin{align}
&\dot{\rho}_{11}= [(\rho_{55}-\rho_{11})\Omega_{15}^2+(\rho_{77}-\rho_{11})\Omega_{17}^2
                  -Re(e^{-2i\phi}\tilde\rho_{13})(\Omega_{15}\Omega_{35}+\Omega_{17}\Omega_{37})]/\Gamma \nonumber\\
&          \quad \quad \quad     +(\rho_{55}+\rho_{66}+\rho_{77})\Gamma/3+(\rho_{22}+\rho_{33}+\rho_{44}-3\rho_{11})\gamma_1, \nonumber\\
&\dot{\rho}_{22}= (\rho_{66}-\rho_{22})\Omega_{37}^2/\Gamma
         +(\rho_{55}+\rho_{66})\Gamma/3+(\rho_{11}-\rho_{22})\gamma_1+(\rho_{33}-\rho_{22})\gamma_z,    \nonumber\\
&\dot{\rho}_{33}= [(\rho_{55}-\rho_{33})\Omega_{35}^2+(\rho_{77}-\rho_{33})\Omega_{37}^2
          +Re(e^{-2i\phi}\tilde\rho_{13})(\Omega_{15}\Omega_{35}+\Omega_{17}\Omega_{37})]/\Gamma \nonumber\\
&          \quad \quad \quad  +(\rho_{55}+\rho_{77})\Gamma/3+(\rho_{11}-\rho_{33})\gamma_1+(\rho_{22}+\rho_{44}-2\rho_{33})\gamma_z, \nonumber\\
&\dot{\rho}_{44}= (\rho_{66}-\rho_{44})\Omega_{35}^2/\Gamma
            +(\rho_{66}+\rho_{77})\Gamma/3+(\rho_{11}-\rho_{44})\gamma_1+(\rho_{33}-\rho_{44})\gamma_z,    \nonumber\\
&\dot{\rho}_{55}=[(\rho_{11}-\rho_{55})\Omega_{15}^2+(\rho_{33}-\rho_{55})\Omega_{35}^2-2Re(e^{-2i\phi}\tilde\rho_{13})\Omega_{15}\Omega_{35}]/\Gamma-\rho_{55}\Gamma\nonumber,\\
&\dot{\rho}_{66}= [(\rho_{22}-\rho_{66})\Omega_{37}^2+(\rho_{44}-\rho_{66})\Omega_{35}^2]/\Gamma-\rho_{66}\Gamma, \nonumber\\
&\dot{\rho}_{77}=[(\rho_{11}-\rho_{77})\Omega_{17}^2+(\rho_{33}-\rho_{77})\Omega_{37}^2-2Re(e^{-2i\phi}\tilde\rho_{13})\Omega_{17}\Omega_{37}]/\Gamma-\rho_{77}\Gamma\nonumber,\\
&\dot{\tilde\rho}_{13}=[e^{2i\phi}(\rho_{11}+\rho_{33}-2\rho_{55})\Omega_{15}\Omega_{35}
                +e^{2i\phi}(\rho_{11}+\rho_{33}-2\rho_{77})\Omega_{17}\Omega_{37} \nonumber\\
&             \quad \quad \quad -\tilde\rho_{13}(\Omega_{15}^2+\Omega_{17}^2+\Omega_{35}^2+\Omega_{37}^2)]/\Gamma/2-(i\Delta+\gamma_2)\tilde\rho_{13}, \nonumber\\
&\frac{\partial \Omega_{15}}{\partial z}+\frac{1}{c}\frac{\partial \Omega_{15}}{\partial t}=
         \alpha [\Omega_{15}(\rho_{55}-\rho_{11})+Re(e^{-2i\phi}\tilde\rho_{13})\Omega_{35}]/\Gamma, \nonumber\\
&\frac{\partial \Omega_{35}}{\partial z}+\frac{1}{c}\frac{\partial \Omega_{35}}{\partial t}=
          \alpha [\Omega_{35}(\rho_{55}+\rho_{66}-\rho_{33}-\rho_{44})+Re(e^{-2i\phi}\tilde\rho_{13})\Omega_{15}]/\Gamma, \nonumber\\
&\frac{\partial \Omega_{17}}{\partial z}+\frac{1}{c}\frac{\partial \Omega_{17}}{\partial t}=
          \alpha [\Omega_{17}(\rho_{77}-\rho_{11})+Re(e^{-2i\phi}\tilde\rho_{13})\Omega_{37}]/\Gamma, \nonumber\\
&\frac{\partial \Omega_{37}}{\partial z}+\frac{1}{c}\frac{\partial \Omega_{37}}{\partial t}=
          \alpha [\Omega_{37}(\rho_{66}+\rho_{77}-\rho_{22}-\rho_{33})+Re(e^{-2i\phi}\tilde\rho_{13})\Omega_{17}]/\Gamma.
\end{align}
\label{Maxwell-Liouville-2}
\end{widetext}

The initial conditions are:
\begin{align}
&\rho_{11}(0,z)=\rho_{22}(0,z)=\rho_{33}(0,z)=\rho_{44}(0,z)=0.25,\nonumber\\
&\rho_{55}(0,z)=\rho_{66}(0,z)=\rho_{77}(0,z)=\tilde\rho_{13}(0,z)=0,\nonumber\\
&\Omega_{15}(0,z)=f_3(0)\cos[\varphi(0)]\Omega,\nonumber\\
&\Omega_{35}(0,z)=f_3(0)\cos[\varphi(0)]\Omega,\nonumber\\
&\Omega_{17}(0,z)=f_3(0)\sin[\varphi(0)]\Omega,\nonumber\\
&\Omega_{37}(0,z)=-f_3(0)\sin[\varphi(0)]\Omega.\
\end{align}
and the boundary conditions:
\begin{align}
&\rho_{11}(t,0)=\rho_{22}(t,0)=\rho_{33}(t,0)=\rho_{44}(t,0)=0.25,\nonumber\\
&\rho_{55}(t,0)=\rho_{66}(t,0)=\rho_{77}(t,0)=\tilde\rho_{13}(t,0)=0,\nonumber\\
&\rho_{11}(t,L)=\rho_{22}(t,L)=\rho_{33}(t,L)=\rho_{44}(t,L)=0.25,\nonumber\\
&\rho_{55}(t,L)=\rho_{66}(t,L)=\rho_{77}(t,L)=\tilde\rho_{13}(t,L)=0,\nonumber\\
&\Omega_{15}(t,0)=f_3(t)\cos[\varphi(t)]\Omega,\nonumber\\
&\Omega_{35}(t,0)=f_3(t)\cos[\varphi(t)]\Omega,\nonumber\\
&\Omega_{17}(t,0)=f_3(t)\sin[\varphi(t)]\Omega,\nonumber\\
&\Omega_{37}(t,0)=-f_3(t)\sin[\varphi(t)]\Omega.
\end{align}
where $f_3(t)$ is the laser pulse envelope generated by the AOM.
In order to take into account rising and falling times, the numerical computation
is performed using for $f_1(t)$, $f_2(t)$ and $f_3(t)$ composite functions,
made of a square wave function between $0$ and $1$, with rising and falling
edges replaced by a sine function during a half period of \SI{50}{\nano\second} for  $f_1(t)$,
\SI{2.5}{\micro\second} for $f_2(t)$ and \SI{0.5}{\micro\second} for $f_3(t)$.


\begin{thebibliography}{9}\label{sec:TeXbooks}


\bibitem{Happer97} T. G. Walker, and W. Happer, Rev. Mod. Phys., {\bf 69}, 629 (1997).

\bibitem{Wynands99} R. Wynands, A. Nagel, Appl. Phys. B., {\bf 68}, 1 (1999).

\bibitem{Vanier05} J. Vanier, Appl. Phys. B., {\bf 81}, 421 (2005).

\bibitem{Arimondo96} E. Arimondo, Prog. Opt. \bf{35}, \rm 257 (1996).

\bibitem{Shah10}	V. Shah, and J. Kitching, Adv. At; Mol. Opt. Phys., {\bf 59},21 (2010).

\bibitem{Happer72} W. Happer, Rev. Mod. Phys., {\bf 44}, 169 (1972).

\bibitem{Taichenachev06} A. V. Taichenachev, V. I. Yudin, V. L. Velichansky, A. S. Zibrov, and S. A. Zibrov, Phys. Rev. A {\bf 73}, 013812 (2006).

\bibitem{Yun14} P. Yun, J-M. Danet, D. Holleville, E. de Clercq, and S. Gu\'erandel, Appl. Phys. Lett., {\bf 105}, 231106 (2014).

\bibitem{Zibrov06}  S. A. Zibrov, V. L. Velichansky, A. S. Zibrov, A. V. Taichenachev, and V. I. Yudin, Opt. Lett. {\bf 31}, 2060 (2006).

\bibitem{Jau04}	Y. Y. Jau, E. Miron, A. B. Post, N. N. Kuzma and W. Happer, Phys. Rev. Lett., {\bf 93}, 160802 (2004).

\bibitem{Shah06}	V. Shah, S. Knappe, P. D. D. Schwindt V. Gerginov, and J. Kitching, Opt. Lett., {\bf 31}, 2335 (2006).

\bibitem{Liu13}	X. Liu, J.-M. M\'erolla, S. Gu\'erandel, C. Gorecki, E. de Clercq, and R. Boudot, Phys. Rev. A {\bf 87}, 013416 (2013).

\bibitem{Zanon05b}	T. Zanon,  S. Gu\'erandel,  E. de Clercq, D. Holleville,  N. Dimarcq, and A. Clairon,Phys. Rev. Lett., {\bf94}, 193002 (2005).

\bibitem{Huang12} M. Huang and J. C. Camparo, Phys. Rev. A, {\bf 85},  012509 (2012).
\bibitem{Godone02}A. Godone, F. Levi, S. Micalizio, and J. Vanier, Eur. Phys. J. D {\bf 18}, 5-13 (2002).


\bibitem{Steck}	D. A. Steck, "Cesium D Line Data," available online at \url{http://steck.us/alkalidata}, (revision 2.1.4, 23 December 2010).

\bibitem{Zanon05}T. Zanon, S. Tr\'emine, S. Gu\'erandel, F. Dah\'es and E. de Clercq, A. Clairon and N. Dimarcq, IEEE Trans. Instrum. Meas., {\bf 54}, 776 (2005).

\bibitem{Park04} S. J. Park, H. Cho, T. Y. Kwon, H. S. Lee,  Phys. Rev. A, {\bf 69},  023806 (2004).

\bibitem{Kozlova12} O. Kozlova, Th\`ese de doctorat, Universit\'e Pierre et Marie Curie, UPMC, Paris, 2012.

\bibitem{Guerandel07}S. Gu\'erandel, T. Zanon, N. Castagna, F. Dah\'es, E. de Clercq, N. Dimarcq, and  A. Clairon, IEEE Trans. Instrum. Meas., {\bf 56}, 383 (2007).

\bibitem{Liu13b} X. Liu, J.-M. M\'erolla, S. Gu\'erandel,E. de Clercq, and R. Boudot,Optics Express, {\bf 21},12451 (2013)

\bibitem{Affolderbach02} C. Affolderbach, S. Knappe, R. Wynands, A. V. Ta.chenachev and V. I. Yudin, Phys. Rev. A {\bf 65}, 043810 (2002).

\bibitem{Taichenachev04} A. V. Taichenachev, V. I. Yudin, V. L. Velichansky, S. V. Kargapoltsev,
R. Wynands, J. Kitching, and L. Hollberg, JETP Letters {\bf 80}, 265 (2004).



\bibitem{Yim08}	S. H. Yim, T. H. Yoon and D. Cho, Rev. Sci. Instrum. {\bf 79}, 126104 (2008).

\bibitem{Scully} M. 0. Scully, M. S. Zubairy, \textit{Quantum Optics}, (Cambridge University Press, Cambridge, 1997).

\bibitem{Yun15} P. Yun, S. Mejri, F. Tricot, M. A. Hafiz, R. Boudot, E. de Clercq, and S. Gu\'erandel, to be published in the proceedings of 8th Symposium on Frequency Standards and Metrology, Potsdam, 2015.

\end{thebibliography}
\end{document}